# Graphene surpasses GaAs/AlGaAs

# for the application of the quantum Hall effect in metrology

Graphene technology demonstrates sufficient maturity to realize a quantum Hall resistance standard able to operate in user-friendly conditions and to confirm the QHE universality, two important contributions to the SI evolution.

R. Ribeiro-Palau[1], F. Lafont[1], J. Brun-Picard[1], D. Kazazis[2], A. Michon[3], F. Cheynis[4], O. Couturaud[5], C. Consejo[5], B. Jouault[5], W. Poirier[1] and F. Schopfer[1*]

[1]LNE - Laboratoire National de Métrologie et d'Essais, avenue Roger Hennequin, 78197 Trappes, France.

[2]LPN - Laboratoire de Photonique et de Nanostructures, CNRS, Route de Nozay, 91460 Marcoussis, France.

[3]CRHEA - Centre de Recherche sur l'Hétéroépitaxie et ses Applications, CNRS, rue Bernard Grégory, 06560 Valbonne, France.

[4]CINAM - Centre interdisciplinaire de nanoscience de Marseille, Campus de Luminy, 13288 Marseille, France.

[5]L2C - Laboratoire Charles Coulomb, CNRS-Université de Montpellier II, Place Eugène Bataillon, 34095 Montpellier, France.

*Correspondence to: felicien.schopfer@lne.fr

**The quantum Hall effect (QHE) theoretically provides a universal standard of electrical resistance in terms of the Planck constant *h* and the electron charge *e*. In graphene, the spacing between the lowest discrete energy levels occupied by the charge carriers under magnetic field is exceptionally large. This is promising for a quantum Hall resistance standard more practical in graphene than in the GaAs/AlGaAs devices currently used in national metrology institutes. Here, we demonstrate that large QHE devices, made of high quality graphene grown by propane/hydrogen chemical vapour deposition on SiC substrates, can surpass state-of-the-art GaAs/AlGaAs devices by considerable margins in their required operational conditions. In particular, in the device presented here, the Hall resistance is accurately quantized within $1\times10^{-9}$ over a 10-T wide range of magnetic field with a remarkable lower bound at 3.5 T, temperatures as high as 10 K, or measurement currents as high as 0.5 mA. These significantly enlarged and relaxed operational conditions, with a very convenient compromise of 5 T, 5.1 K and 50 μA, set the superiority of graphene for this application and for the new generation of versatile and user-friendly quantum standards, compatible with a broader industrial use. We also measured an agreement of the quantized Hall resistance in graphene and GaAs/AlGaAs with an ultimate relative uncertainty of $8.2\times10^{-11}$. This supports the universality of the QHE and its theoretical relation to *h* and *e*, essential for the application in metrology, particularly in view of the forthcoming Système International d'unités (SI) based on fundamental constants of physics, including the redefinition of the kilogram in terms of *h*.**

The quantum Hall effect (QHE) is a macroscopic quantum phenomenon corresponding to the quantization of the charge carrier cyclotron orbits occuring in two-dimensional conductors under magnetic field *(1)*. Here, Heisenberg's uncertainty and Pauli's principles are directly manifested by relating the quantized (transverse) electrical resistance $R_H$ to the electron charge $e$ and Planck's constant $h$ only. The theoretical relationship $R_H=h/(ie^2)$, with i an integer, provides, in principle, a universal and reproducible electrical resistance standard. Moreover, this equation is of crucial importance in the ongoing effort to redefine several units of the *Système International d'unités* (SI) in terms of fundamental constants of physics by exploiting basic physical laws or equivalence between quantities, as it is notably the case for the kilogram in terms of $h$ through the watt balance *(2)*. Important evidence of the exactness of the relation $R_H=h/(ie^2)$ comes from the tests of the universality of $R_H$, *i.e.* its invariance with the material used to realize the QHE devices *(3)*. Such experiments are also tests of the strong theoretical arguments underpinning the validity of this relation (insensitivity of the quantized Hall resistance, as a topological invariant *(4,5,6)*, to electron-electron interaction, gravity *(7)*) or questioning it (quantum electrodynamics corrections *(8)*). Previous results *(9)* have been used to recommend the QHE as the official representation of the ohm, enabling routine resistance calibrations with accuracy within $1\times10^{-9}$ in national metrology institutes, all over the world *(10)*. Extending these QHE universality tests is of growing interest at the time of the SI redefinition, and deserves to meet the challenge set by such experiments at the limit of instrumentation.

In the context of a broader and better integration of quantum physics in the SI, the simplification of the experimental conditions of the quantum Hall resistance standard (QHRS) is, for metrologists, as much important as the aforementioned fundamental issues. The most widely used QHRS today, based on GaAs/AlGaAs heterostructures (GaAs-QHRS), typically require a magnetic field ($B$) around 10 T, unachievable without expensive superconducting cryomagnetic systems, a temperature around $T$=1.3 K, and measurement currents ($I$) that cannot exceed tens of µA, in order to preserve a $1\times10^{-9}$-accuracy of $R_H$ *(11)*. Such low currents require a specific low temperature amplifier based on a superconducting quantum interference device in order to improve the signal-to-noise ratio and achieve a $10^{-9}$-precision in the calibration experiments against the primary standard. This precision is necessary in order to ensure the final uncertainty required by industrial needs, unavoidably degraded over the resistance traceability chain. Relaxing the operational conditions ($B$, $T$, $I$) to make the QHRS compatible with transportable, compact helium-free setup and commercial amplifiers, will simplify their handling and will reduce the calibration cost, leading to a broader dissemination beyond national metrology institutes, particularly in industry. This simpler and more available quantum resistance standard will improve not only the ohm traceability in DC and AC regime, but also the farad *(12)*, and, more importantly, the ampere, directly in terms of its new definition based on a fixed value of $e$ *(13)*. Beyond these electrical quantities, a user-friendly quantum resistance standard will benefit to the traceability of mass, temperature, and so on, since electrical measurements are ubiquitous. Nevertheless, the QHE physics makes the relaxation of these three experimental parameters ($B$, $T$, $I$) very challenging since they are interdependent: decreasing $B$ competes against increasing $T$ and $I$, while seeking to preserve the device accuracy.

Here we present a QHRS fabricated from graphene, a two-dimensional crystal of carbon atoms, that demonstrates all the advantages. It can operate with a $1\times10^{-9}$-accuracy over a wide range of ($B$, $T$, $I$), exceeding the range of previous graphene devices and, more importantly, of semiconductors devices. New absolute records are established (Fig. 1) for each operational condition, either $B$ or $T$, or $I$, (with the remaining two parameters maintained at workable values, *i.e.* at least similar to what is required by GaAs-QHRS): i) minimum operation magnetic field of 3.5 T, three times smaller than in a usual GaAs-QHRS, ii) range of quantizing magnetic field larger than 10 T, ten times wider than in a usual GaAs-QHRS, iii) maximum operation temperature of 10 K, seven times higher than in a usual GaAs-QHRS, iv) maximum operation current of 0.5 mA, ten times higher than in a usual GaAs-QHRS. We also demonstrate that the device keeps its $1\times10^{-9}$-accuracy under an attractive set of relaxed experimental conditions never simultaneously achieved before: 5 T, 5.1 K, and 50 μA, making a breakthrough in the simplification of the QHRS implementation. Finally, this ideal graphene QHRS (G-QHRS) allowed an improved test of the QHE universality through a comparison with a GaAs-QHRS that showed an agreement within a record relative uncertainty of $8.2\times10^{-11}$.

The ability of graphene to exhibit a robust QHE at low $B$ derives from the fact that charge carriers behave as massless relativistic particles, with a peculiar quantization of the energy spectrum in Landau level (LL) under magnetic field. The energy spacing between the first two degenerated LLs is given by $\Delta E_{G}(B)=36\sqrt{B}\,\mathrm{meVT}^{-1/2}$ *(14),* while it is of $\Delta E_{GaAs}(B)=1.7B\,\mathrm{meVT}^{-1}$ in GaAs devices. Because of this difference, the energy spacing required for a Hall resistance to be accurately quantized at $h/(2e^2)$ within $1\times10^{-9}$ at 1.3 K occurs at $B\approx10$ T for GaAs devices, but only at $B\approx0.2$ T for graphene devices. Under a few teslas, the energy spacing is so large in graphene that operation temperature and current are expected to be higher than in GaAs.

Nevertheless, realizing the expected benefits of graphene in a practical QHRS, which would operate at lower $B$, higher T, and higher $I$, while preserving accuracy within $1\times10^{-9}$, is quite challenging in terms of material quality, as suggested by several unsuccessful attempts *(15, 16, 17, 18, 19, 20)*. This should require large Hall bars (≈10 000 $\mu m^2$, as learned from previous studies of graphene) made of high-quality uniform monolayer with relatively high carrier mobility (≥10 000 $cm^2V^{-1}s^{-1}$) and where the carrier density can be homogeneously set below $2\times10^{11}cm^{-2}$ (if targeting $B\leq4$ T) *(21)*. Low-resistance contacts (~10 Ω) between graphene and metallic pads are also indispensable for precise electrical measurements of the quantum Hall state, as learned from the experience with GaAs-QHRS *(22)*. An encouraging milestone was achieved by Tzalenchuk *et al. (23,24,25),* who demonstrated excellent $R_H$ accuracy down to $8.7\times10^{-11}$ in epitaxial monolayer graphene grown by silicon sublimation from SiC, but at $T$ around 0.3 K and $B$ around 14 T, operation conditions that prevent such graphene device from advantageously replacing GaAs-QHRS.

Our device is a large 100 μm×420 μm graphene Hall bar, as shown in Fig. 2a. Graphene was grown on the Si-face of a semi-insulating 0.16° off-axis 6H-SiC substrate by propane/hydrogen chemical vapor deposition (CVD) *(26)* (see supplementary). This original hybrid method is expected to benefit from the advantages of both the growth on SiC as a monocrystalline substrate *(27)* and CVD as a tunable growth method *(28)*. Figure 2b shows the four-terminal transverse (Hall) resistance $R_H$ and the four-terminal longitudinal resistance $R_{xx}$ (per square), measured at 1.3 K and up to 14 T. At low magnetic field, well-known quantum corrections to $R_{xx}$ are observable (*e.g.* weak-localization and electron-electron interaction *(29)*) and $R_H$ manifests the classical Hall effect. From $B$ = 1 T, the LLs become well separated and we observe a few Shubnikov-de Haas oscillations (SdH) of $R_{xx}$, while $R_H$ starts to mark a plateau, around $h/(6e^2)$, characteristic of graphene. The Hall coefficient gives an electron density of $n_s\sim1.8\times10^{11}$ $cm^{-2}$, which turned out to be remarkably homogeneous (within $1.5\times10^{10}$ $cm^{-2}$) over the large surface of the device. The extracted carrier mobility is $\mu$~9 400 $cm^2V^{-1}s^{-1}$. The metallic contacts to graphene were found of outstanding quality with a resistance below 1.2 Ω (see Fig. 2a, and supplementary).

At 1.3 K, the so-called $h/(2e^2)$ quantum Hall resistance plateau, corresponding to LLs filling factor expected to be around $\nu=n_s h/(eB)=2$, is observable from 2.5 T and remains flat up to at least 14 T, the highest *B* available in our setup (Fig. 2b). Over this *B*-field range, $R_{xx}$ has a very low value (as described below), another direct manifestation of the quantum Hall state where charge transport is expected to occur without dissipation at zero temperature. For comparison, in a commonly used GaAs-QHRS (LEP514) *(30)*, the $h/(2e^2)$ plateau, centred at 10.8 T, extends only over 2 T, at 1.3 K (Fig. 2b). The large extension of the Hall resistance plateau observed in our device evidences the robustness of the QHE in graphene. An even more spectacular manifestation of this robustness, is the persistence of the wide $h/(2e^2)$ plateau up to 100 K, from 8 T with an accuracy within a few $10^{-3}$. This advantageously compares with the previous remarkable observation of $h/(2e^2)$ Hall resistance plateaus with similar accuracy at 300 K, but 45 T in ref *(31)* and demonstrates the progress achieved in the graphene quality at lower carrier density.

The accuracy of $R_H$ in the G-QHRS, denoted $R_{H\text{-}G}$, was extensively studied on the $h/(2e^2)$ plateau at low temperature, through indirect comparisons of $R_{H\text{-}G}$ with the resistance of a well characterized GaAs-QHRS (LEP514), $R_{H\text{-}GaAs}$, *via* an intermediate 100-Ω room-temperature resistor, using a cryogenic current comparator bridge (see supplementary). Figure 3a shows the relative deviation between the two devices, $(\Delta R_H/R_H)_{\text{-}G}=(R_{H\text{-}G}-R_{H\text{-}GaAs})/R_{H\text{-}GaAs}$, which can be used to characterize the quantization accuracy of $R_{H\text{-}G}$. In the *B*-range 3.5 T-14 T, $R_{H\text{-}G}$ is accurately quantized with no significant deviation within the relative combined standard uncertainty of the measurements $u_c$, equal to $1\times10^{-9}$. This quantized Hall resistance plateau, more than 10-T wide, is remarkable: in GaAs-QHRS the $1\times10^{-9}$-accuracy of $R_{H\text{-}GaAs}$ typically does not spread over much more than 1 T (Fig. 2b). Attempts to explain such a wide quantized plateau involves two mechanisms: a charge transfer from the substrate stabilizing the Landau level filling factor at $\nu=2$ when increasing the magnetic field, and a certain disorder, both resulting from the special structure of graphene on SiC *(32,33)*. Nevertheless the wide quantized plateau is very convenient by eliminating the adjustment procedure of *B* as required in GaAs-QHRS. Supporting the accurate quantization of the Hall resistance, $R_{xx}$, measured in G-QHRS using a SQUID (see supplementary), displays very low values, below 35 μΩ (Fig. 3b), over the *B*-range 3.5 T-14 T, at 1.3 K, with a minimum of (5±2) μΩ reached around 5 T- 6 T.

Remarkably, at 3.5 T, $(\Delta R_H/R_H)_{\text{-}G}$ is measured equal to $(+0.5\pm1.4)\times10^{-9}$. This accuracy was demonstrated at 1.3 K and using 10 μA, *i.e.* in workable conditions of temperature and current similar to those required by GaAs-QHRS. This operation magnetic field is much lower than the lowest of 11.5 T, achieved till now in graphene devices grown by other techniques *(25)*, while preserving $R_H$ accurately quantized within $1\times10^{-9}$ (Fig. 1b). Importantly, lower than the 10 T magnetic field required for a usual GaAs-QHRS, it also beats the previous record value of 6 T exceptionally reported in a state-of-the-art GaAs-QHRS (with an accuracy not better than a few $10^{-9}$ only) *(11)*. From a practical point of view, this low operation magnetic field is achievable with a permanent magnet *(34)*, which is an asset to develop very practical QHRS setups.

At 5 T, 1.3 K and with 50 μA, we have repeated $(\Delta R_H/R_H)_{\text{-}G}$ measurements (Fig. 6a). Their weighted mean, equal to $(-11.4\pm10.2)\times10^{-11}$, demonstrates the outstanding performance of our graphene device which displays a state-of-the-art accuracy at very low *B*.

The robustness of the Hall resistance quantization was tested at higher temperature (Fig. 4). The critical temperature $T_c$, below which $R_{H\text{-}G}$ stays accurately quantized within $1\times10^{-9}$, increases with the magnetic field (Fig. 4c), reaching a remarkable maximum of 10 K at 8.5 T and 20 μA. This operation temperature is higher than the typical operation temperature of 1.3 K of a usual GaAs-QHRS. Remarkably, it also exceeds the highest similar critical temperature that can be deduced from previous experiments in epitaxial graphene, equal to 7.5 K and obtained at a much higher magnetic field of 14 T *(25)* (Fig 1c). From a practical point of view, the value of 10 K is well above liquid helium temperature and basic specification of common pulse-tube cryocoolers *(35)*.

The robustness of the quantization accuracy at higher *I* is an advantage to improve and simplify the resistance measurement because enabling higher signal to noise ratio. It was mainly tested by $R_{xx}$ measurements (Fig. 5). Assuming a coupling factor between $(\Delta R_H/R_H)_{\text{-}G}$ and $R_{xx}$ with a maximum absolute value of 0.6 (see supplementary), $R_{H\text{-}G}$ should remain quantized within $1\times10^{-9}$ for $R_{xx}$ below 25 μΩ. The critical current, $I_c$, below which $R_{H\text{-}G}$ remains accurately quantized within $1\times10^{-9}$ ($R_{xx} \leq 25$ μΩ) increases with the magnetic field reaching a remarkable maximum of 0.5 mA at 8 T and 1.3 K (Fig. 5b). Fig. 5b, also presents $I_c(B)$ measured at 5 K, which reaches a maximum of 0.27 mA at 8 T. The high critical current, achieved at relatively low magnetic field at 1.3 K, completes the figures of merit of the G-QHRS presented. It is much higher than the operation current of a usual GaAs-QHRS. It is comparable to the record value of 0.6 mA exceptionally achieved in a state-of-the-art GaAs-QHRS at 10 T and 1.3 K but of width (1 mm) ten times larger than that of our G-QHRS *(16)*. This highlights the much higher current density sustained in graphene, as already evidenced in previous experiments *(25)* by the achievement of 0.4 mA in a 35-μm wide device, at 0.3 K and 14 T (Fig. 1b).

Searching for the most interesting simultaneous operation conditions (*B*, *T*, *I*) ensuring $1\times10^{-9}$-accuracy of this device inside its wide parameter space, it appears that (5 T, 5.1 K, 50 μA) constitutes a very convenient combination of parameters that were inaccessible up to now for a QHRS, Fig. 1e. Such operation conditions enable a real breakthrough in the instrumentation for the implementation of quantum Hall resistance standard.

As proofs of the graphene device quality, we first mention the excellent spatial homogeneity, over large scale, as established by the agreement within $1.5\times10^{-9}$ of the $R_H$ measurements performed using the three transversal terminal-pairs of the device over the wide *B*-range 4 T - 10.8 T (see supplementary). This can be explained, not only by the homogeneity of the graphene morphology at long length scale, as imaged by low energy electrons microscopy (LEEM) (see supplementary) but also by the quality of the device fabrication especially that of metallic contacts. In addition, all the electronic transport properties, including $n_s$, μ, contact resistances, as well as the Hall quantization ($R_{xx}$, $R_{H\text{-}G}$) are reproducible, without any ageing nor degradation, over, so far, 10 months (device kept under helium atmosphere or vacuum) and after tens of thermal cycling. The far better performances of this graphene-based QHRS compared to a previously produced one, of slightly higher carrier density $n_s$~$3.2\times10^{11}$ cm$^{-2}$, lower mobility $\mu$~3 500 cm$^2$V$^{-1}$s$^{-1}$ and showing a $1\times10^{-9}$-accuracy above 10 T *(33)* demonstrates not only the reproducibility but also the progress yet accomplished of the graphene device fabrication process based on CVD on SiC. LEEM images of the graphene layer used for the fabrication of the most demonstrative device show residual small sub-micrometric inclusions of bilayer graphene and suggests there is room for further improvement of the quality of the graphene device produced by the proposed method.

Interestingly, the measurement of a robust $1\times10^{-10}$-accurate Hall quantization in presence of these bilayer inclusions confirm the strong immunity of the QHE against such short-range defects (compared to the Hall bar width).

Finally, considering the high quality of our G-QHRS, we used it to perform a QHE universality test, at the best level, by comparing $R_{\mathrm{H}}$ in the graphene device and in a GaAs-QHRS. Twenty-eight measurements carried out at 1.3 K, between 4 T and 8 T with the majority at 5 T (Fig. 6a), were selected using severe criteria based on longitudinal and Hall resistance values (see supplementary). For each measurement, the combined standard uncertainty $u_{\mathrm{c}}$ is dominated by uncorrelated random noise as demonstrated by the Gaussian description of the set of measured values and by the Allan deviation calculations *(36)* (Fig. 6a and supplementary). The weighted mean of these measurements shows an agreement between $R_{\mathrm{H\text{-}G}}$ and $R_{\mathrm{H\text{-}GaAs}}$ to $-0.9\times10^{-11}$ within $u_{\mathrm{c}}=8.2\times10^{-11}$. This extremely precise comparison of quantized Hall resistances in graphene and GaAs/AlGaAs displays a record uncertainty in the field of these very difficult QHE universality tests, better than that of $8.7\times10^{-11}$ obtained in a previous similar experiment by Janssen *et al. (25)*. Besides, our experiment is performed with graphene produced by CVD on SiC, a technique different from that of silicon sublimation from SiC used in the experiment by Janssen *et al.* Considering both results gives a proof of the reproducibility of the quantized Hall resistance in graphene independently of the fabrication technique. Then they can be combined to establish the new state-of-the-art of the QHE universality tests, which shows an agreement between $R_{\mathrm{H\text{-}G}}$ and $R_{\mathrm{H\text{-}GaAs}}$ to $-2.7\times10^{-11}$ within $u_{\mathrm{c}}=6.0\times10^{-11}$ (Fig. 6b). Our new result, proving the invariance of $R_{\mathrm{H}}$ within such a low uncertainty in materials as different as graphene and GaAs/AlGaAs, in terms of structure, disorder and physics, gives an additional strong support to the exactness of the QHE relation $R_{\mathrm{H}}=h/(\mathrm{i}e^2)$, *i.e.* of the QHE theory itself. This is an important result for fundamental metrology. Indeed, the QHE relationship can be combined with the Josephson effect relationship, used to realize an electrical voltage standard in terms of $h/(2e)$, in order to establish a direct link between the unit of mass and $h$. Experimentally, this is achieved by the calibration of a mechanical power in terms of an electrical power in the watt balance experiment *(37)*. This allows to consider a new definition of the unit of mass, more universal that would advantageously replace the international prototype of the kilogram dated from 1889 kept in a vault near Paris, and slowly changing in mass with time. Figure 6b sums up the most salient QHE comparisons carried out between GaAs/AlGaAs and Si-Mosfet *(22,38)*, on one hand, and between GaAs/AlGaAs and graphene, on the other hand *(25)*. It shows, over time, an improvement of the accuracy, together with a simplification of the experimental conditions, allowed by graphene. This proves the continuous progress in the control of the quantum Hall state, that also bodes well for the new SI.

In conclusion, we report on the measurement of an ideal and versatile graphene quantum Hall resistance standard, which operates over a wide range of experimental conditions while keeping accuracies within $1\times10^{-9}$, and possibly as low as $8.2\times10^{-11}$. Remarkably, it can operate at very low magnetic field (3.5 T) or high temperature (10 K), or high current (0.5 mA), establishing new records of relaxed operational conditions for a QHRS. Hence, we demonstrate, ten years after the first measurements of its exceptional electronic properties, that graphene realizes its promises and outperforms other well-studied semiconductors, in the very demanding resistance metrology application. The abandonment of GaAs/AlGaAs for this application is inevitable. The demonstrated performances rely on the quality, large scale homogeneity, and reproducibility of the graphene grown by CVD on SiC. The demonstrated maturity of the material is promising for the next success of graphene in other applications *(39)*. The convenient G-QHRS developed opens the era of easy-handling, helium free, and affordable quantum standards able to broadly disseminate, towards industrial end-users, the units defined in the new SI from the fundamental constants of physics. In addition, considering the new demonstration of the QHE universality with ultimate precision and its implication in the kilogram redefinition, does make graphene further weighting in the evolution of the SI.

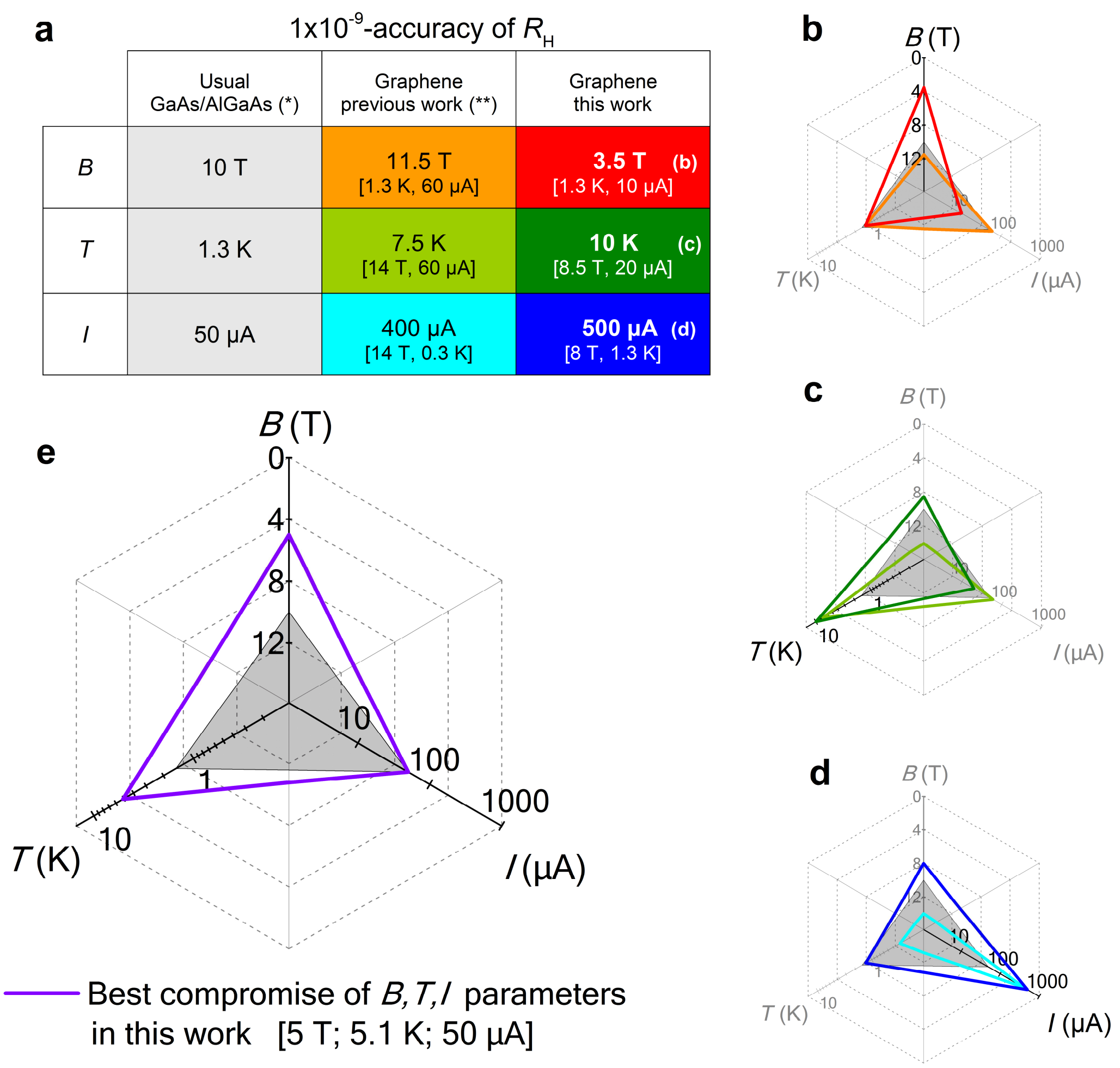

**a** $1\times10^{-9}$-accuracy of $R_H$

| | Usual GaAs/AlGaAs (*) | Graphene previous work (**) | Graphene this work |
|---|---|---|---|
| *B* | 10 T | 11.5 T [1.3 K, 60 μA] | **3.5 T** **(b)** [1.3 K, 10 μA] |
| *T* | 1.3 K | 7.5 K [14 T, 60 μA] | **10 K** **(c)** [8.5 T, 20 μA] |
| *I* | 50 μA | 400 μA [14 T, 0.3 K] | **500 μA** **(d)** [8 T, 1.3 K] |

**Fig. 1. Operational conditions of magnetic field *B*, temperature *T* and measurement current *I* ensuring accurate quantization of the Hall resistance within $1\times10^{-9}$. (a)** New absolute records established by our G-QHRS, compared to performances previously demonstrated in epitaxial graphene devices, and in usual GaAs-QHRS. (*) Ref. 11 and (**) Ref. 25. The conditions of the records in *B*, *T*, and *I* are illustrated in diagrams **(b)**, **(c)**, and **(d)** respectively. **(e)** Remarkable set of (*B*,*T*,*I*), which together enables our G-QHRS to operate.

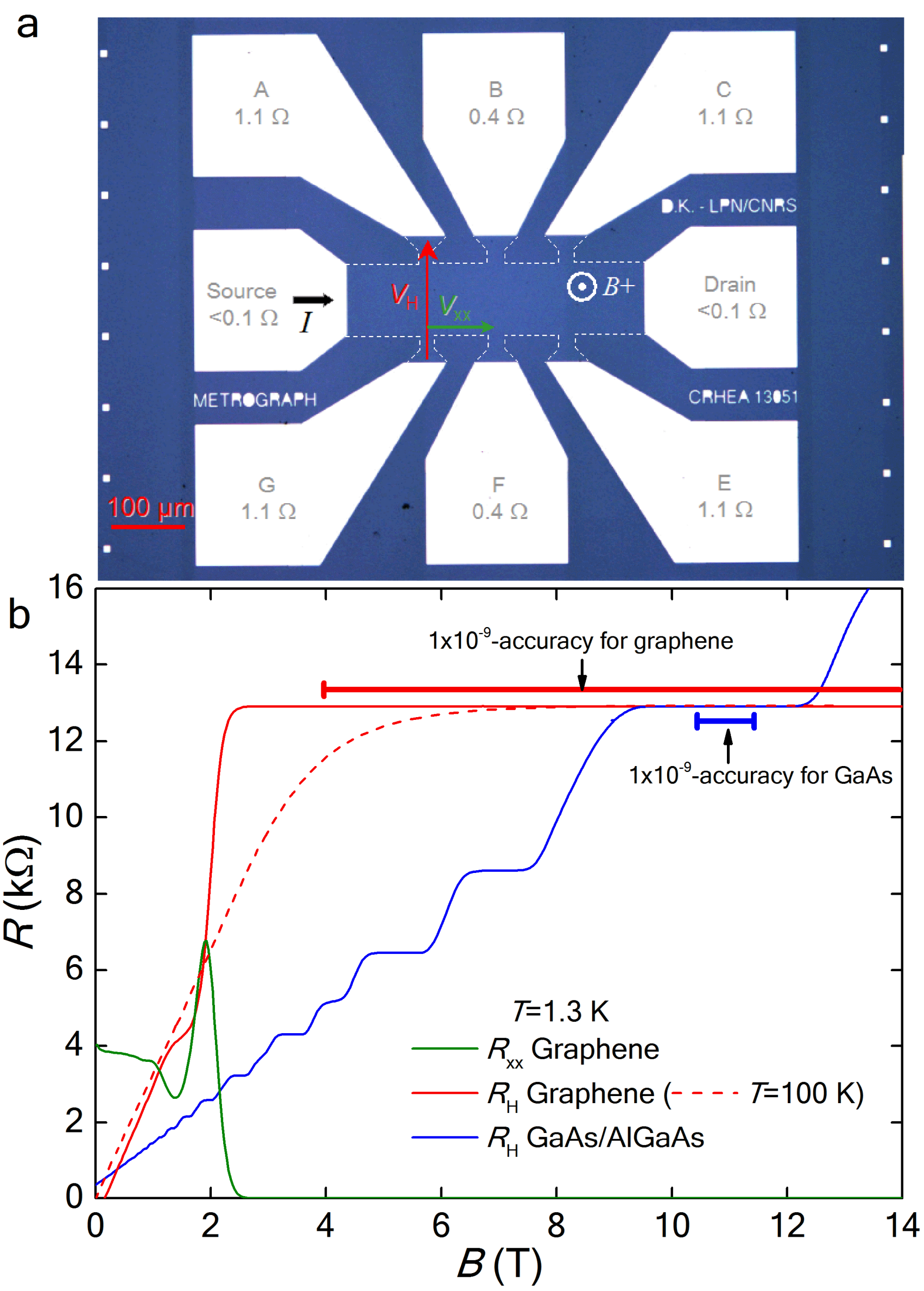

**Fig. 2. Device and magneto-resistance. (a)** Optical image of the measured G-QHRS. The resistance of each Ti/Pd/Au contact is indicated. (**b**) Longitudinal resistance per square (green) and Hall resistance (red) for the graphene device and Hall resistance (blue) for GaAs/AlGaAs device, versus magnetic field, at 1.3 K, with a current $I$=0.1 µA. $R_{xx}$ versus magnetic field for the graphene device at 100 K (Red dashed line). Red (blue) horizontal line represents the magnetic field interval where the graphene (GaAs/AlGaAs) device presents a $1\times10^{-9}$ accuracy on $R_H$. Generically, the longitudinal resistance per square is $R_{xx}=(w/l)V_{xx}/I$ where $l$ is the length over which $V_{xx}$ is measured, $w$ is the Hall bar width, and the Hall resistance is $R_H = V_H/I$, as exemplified on the device picture.

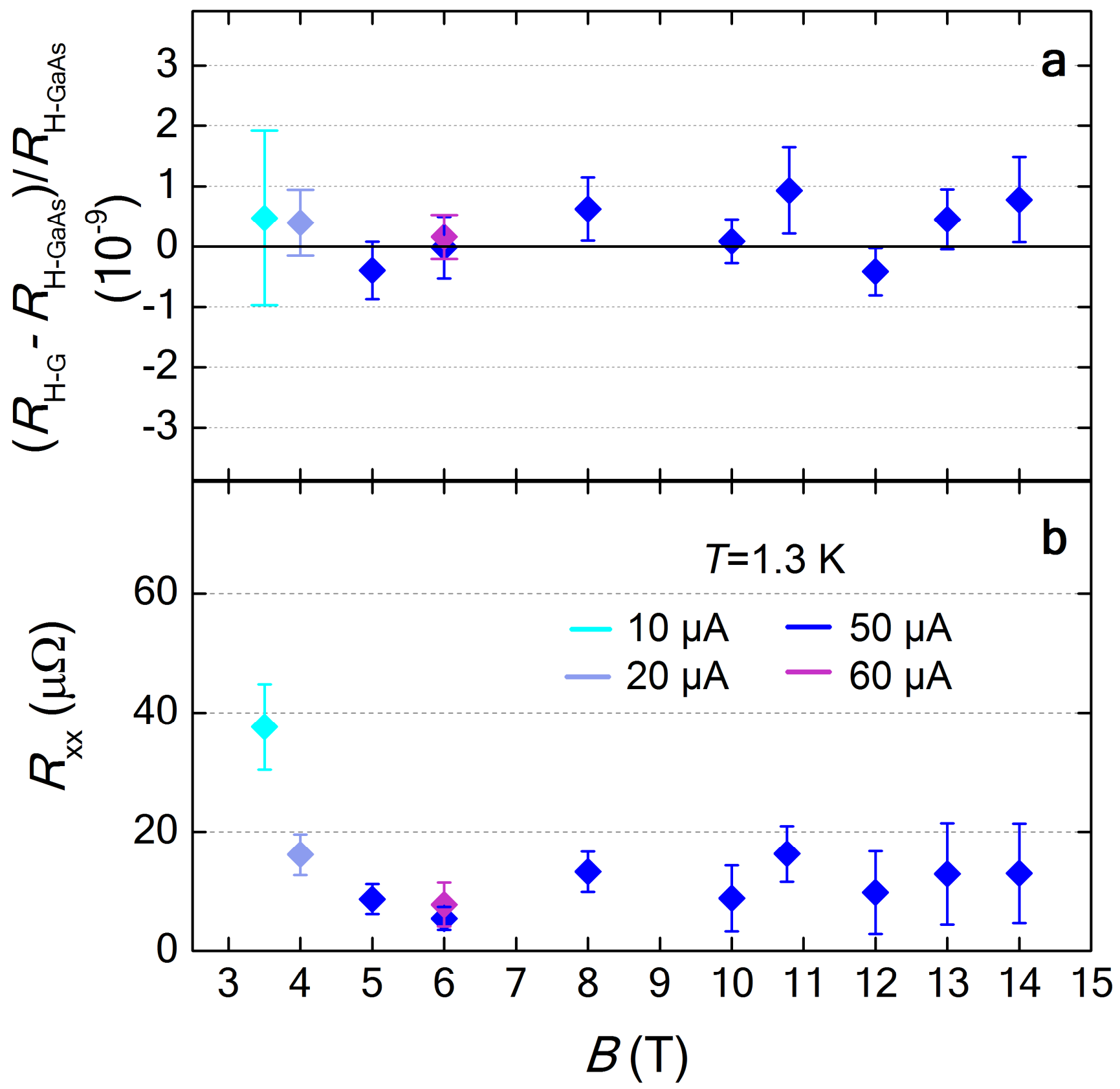

**Fig. 3. Precision measurements of the Hall and longitudinal resistance of the G-QHRS along the $h/(2e^2)$ plateau, at 1.3K .** (**a**) Relative deviation of $R_{H-G}$ from $R_{H-GaAs}$, measured using the central terminals (B,F) of the graphene device, and (**b**) mean value of the four measurements of the longitudinal resistance per square along the device edges, versus magnetic field. The operation currents are indicated.

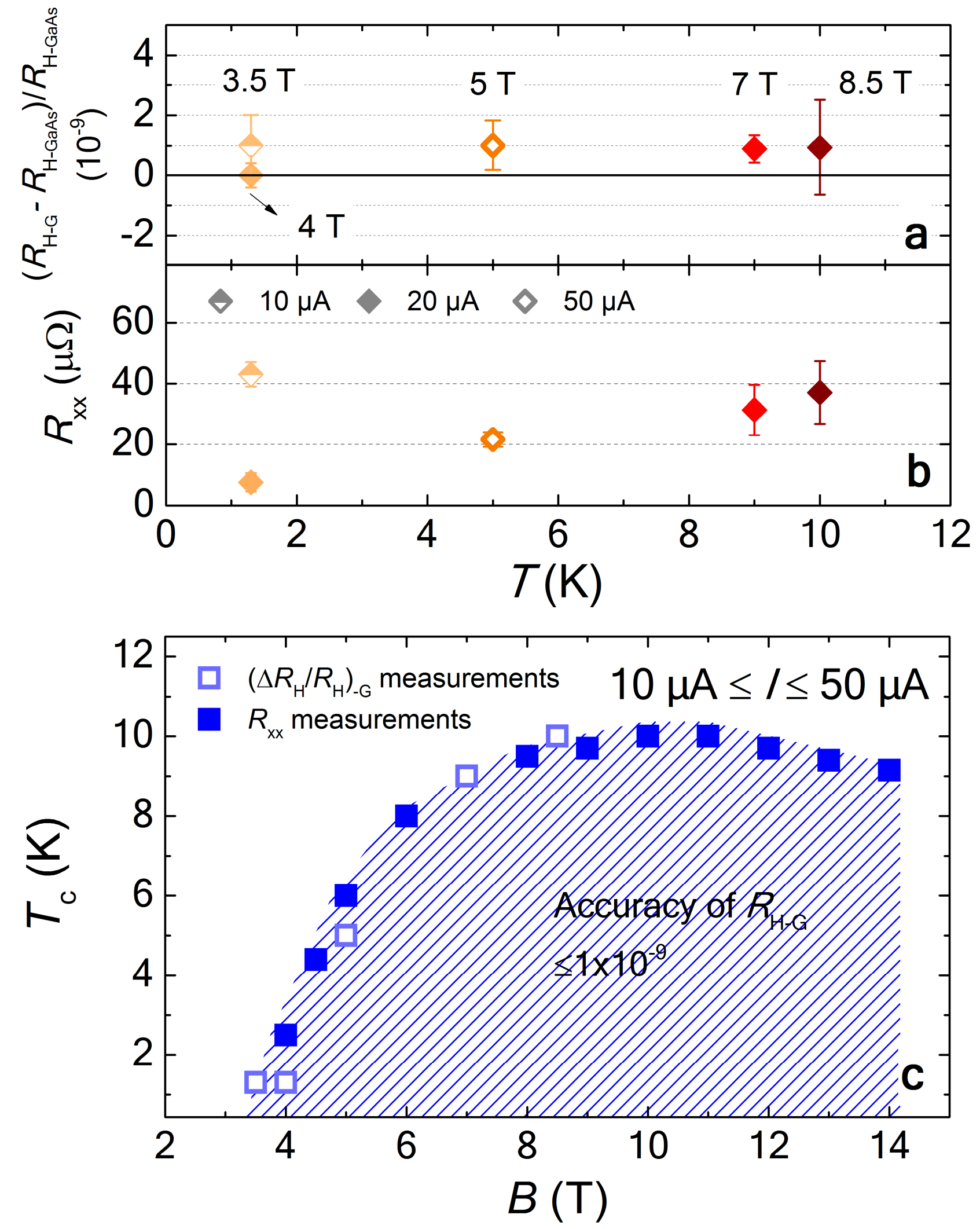

**Fig. 4 Precision measurements of the Hall and longitudinal resistance of the G-QHRS along the $h/(2e^2)$ plateau, at increasing temperature.** (**a**) Relative deviation of $R_{H\text{-}G}$ from $R_{H\text{-}GaAs}$, measured using the central terminals (B,F) of the graphene device, and (**b**) mean value of the four measurements of the longitudinal resistance per square along the device edges, versus temperature. The operation magnetic fields and currents are indicated. **(c)** Critical temperature, defined as the temperature below which $R_{H\text{-}G}$ is accurately quantized within $1\times10^{-9}$, deduced from $R_{xx}$ and $(\Delta R_H/R_H)_{\text{-}G}$ measurements presented in (a) and (b) (filled symbols), and from $R_{xx}$ measurements only (using the criterion $R_{xx} \leq 25\ \mu\Omega$) (open symbols). $T_c$ is plotted as a function of the magnetic field for a given range of operation current between 10 µA and 50 µA.

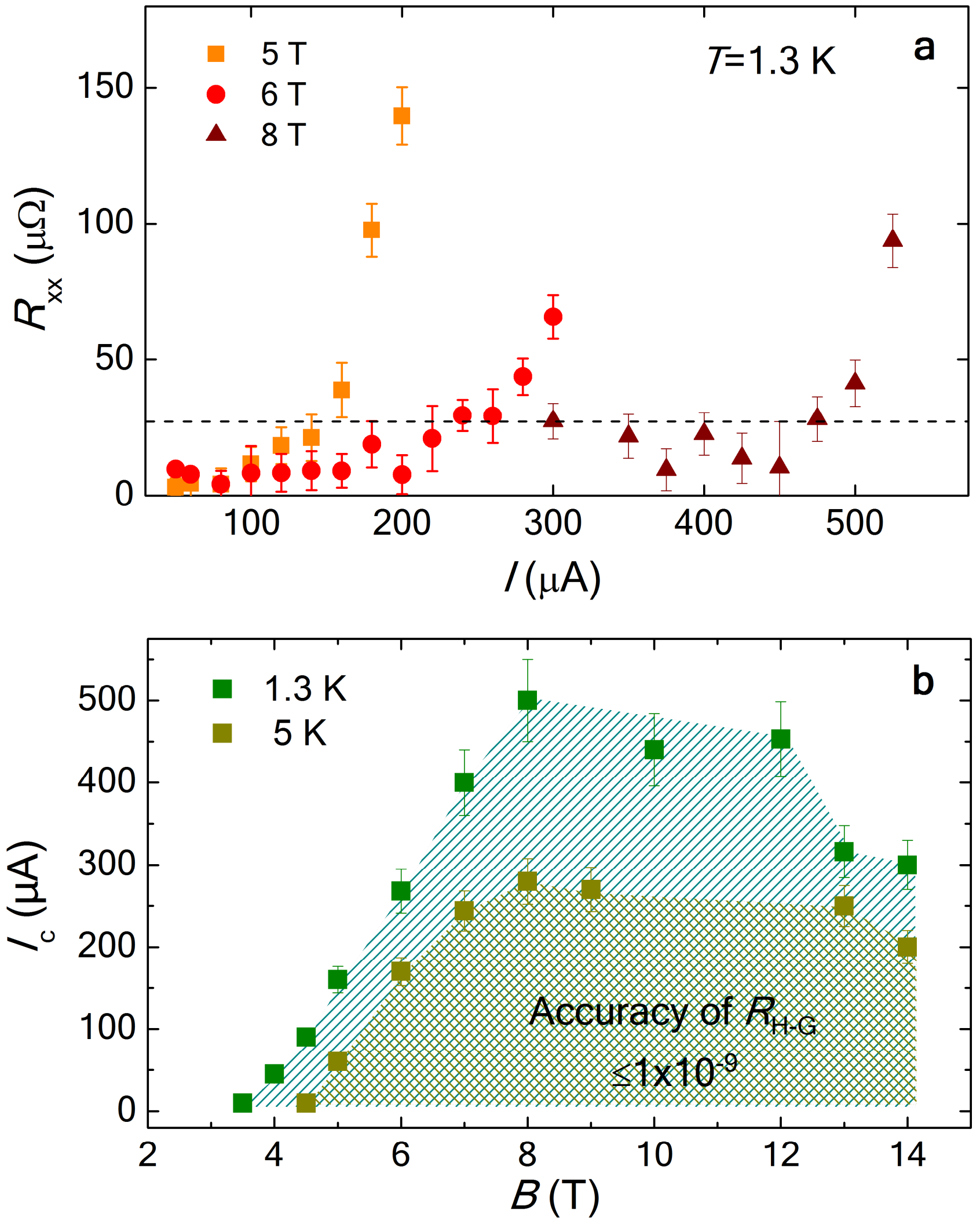

**Fig. 5 Precision measurements of the longitudinal resistance of the G-QHRS along the *h*/(2*e*$^2$) plateau, at increasing current.** (**a**) Mean value of the longitudinal resistance per square versus operation current at 5 T, 6 T, and 8 T, at 1.3 K. **(b)** Critical current, defined as the operation current below which $R_{H-G}$ stays accurately quantized within 1×10$^{-9}$, deduced from $R_{xx}$ measurements similar to those presented in (a). $I_c$ is plotted as a function of the magnetic field, for *T*=1.3 K and for *T*=5 K. Error bars correspond to the dispersion of the $I_c$ values deduced from repeated measurements. The significant dispersion observed for the highest $I_c$ values can be explained by instability manifested when approaching the breakdown of the quantum Hall state at very high Hall voltage drops (≈5 V).

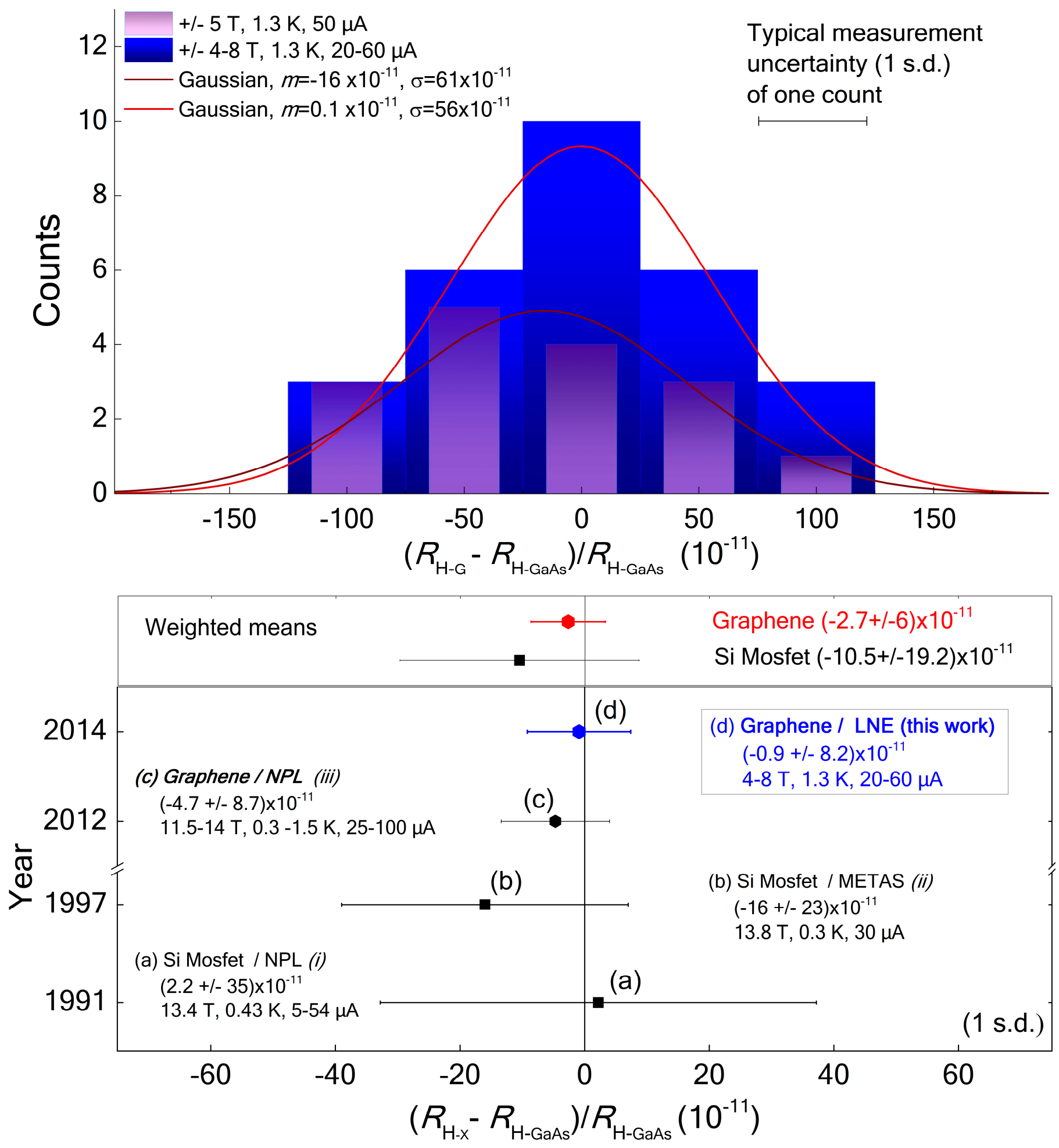

**Fig. 6. QHE universality test.** (**a**) Histograms of the relative deviations of $R_{\text{H-G}}$ from $R_{\text{H-GaAs}}$, measured with graphene at (±5 T, 1.3 K, 50 μA) (violet), and of all the values in the range ±(4–8) T, at 1.3 K with currents 20–60 μA (blue), including the previously mentioned subset. Wine and red curves are Gaussian evaluated using the statistical analysis (mean $m$ and standard deviation σ) of the corresponding series of measurements. The weighted mean of the measurements is $(-11.4\pm10.2)\times10^{-11}$ (violet) and $(-0.9\pm8.2)\times10^{-11}$ (blue). (**b**) Most salient QHE comparisons between different materials over the time, including this work. Top inset: weighted means of the quantized Hall resistance comparisons in Si Mosfet and GaAs/AlGaAs (measurements (a) and (b)), and in Graphene and GaAs/AlGaAs (measurements (c) and (d)). (i) ref 38, (ii) ref 22 and (iii) ref 25.

**Acknowledgments** We wish to acknowledge C. Berger (GeorgiaTech), S. Borini (Nokia Research center), D. Estève (CEA), D.C. Glattli (CEA), P. Gournay (BIPM), M. Keller (NIST), and K. von Klitzing (Max Planck Institute) for helpful discussions, D. Leprat (LNE), D. Mailly (LPN), M. Portail (CRHEA), T. Chassagne (NovaSiC), M. Zielinski (NovaSiC) for technical support. This research has received funding from the french Agence Nationale de la Recherche (Grant No. ANR-2011-NANO-004). It has been partly supported within the European Metrology Research Programme (EMRP) project SIB51, GraphOhm. The EMRP is jointly funded by the EMRP participating countries within the European Association of National Metrology Institutes (EURAMET) and the European Union.

**Author contributions** F. S., R. R.-P., W. P. planned the experiments. A. M. fabricated the graphene layers. F. C., A. M. performed structural characterization of the graphene layers. D. K. fabricated the Hall bar devices. R. R-P., F. S., J. B.-P., W. P., F. L. conducted the electrical metrological measurements. O. C., C. C., B. J. and D. K. carried out complementary electrical measurements. R. R-P., F. S., W. P. analyzed the data. F.S., R. R.-P., W. P., F. L., B. J., A. M., D. K. wrote the paper.

---

[21] Schopfer, F. and Poirier, W. Graphene-based quantum Hall effect metrology. *MRS bulletin* **37,** 1255 (2012).

[22] Jeckelmann, B., Jeanneret, B., Inglis, D. High-precision measurements of the quantized Hall resistance: experimental conditions for universality. *Phys. Rev. B* **55,** 13124 (1997).

[23] Tzalenchuk, A. *et al.* Towards a quantum resistance standard based on epitaxial graphene. *Nature Nanotech*. **5,** 186 (2010).

[24] Janssen, T. J. B. M. *et al.* Graphene, universality of the quantum Hall effect and redefinition of the SI system. *New J. Phys.* **13,** 093026 (2011).

[25] Janssen, T. J. B. M. *et al.* Precision comparison of the quantum Hall effect in graphene and gallium arsenide *Metrologia* **49,** 294 (2012).

[26] Michon, A. *et al.* Direct growth of few-layer graphene on 6H-SiC and 3C-SiC/Si via propane chemical vapor deposition. *Appl. Phys. Lett.* **97,** 171909 (2010).

[27] Berger, C. *et al.* Electronic confinement and coherence in patterned epitaxial graphene. *Science* **3012,** 1191 (2006).

[28] Michon, A. *et al.* Effects of pressure, temperature, and hydrogen during graphene growth on SiC(0001) using propane-hydrogen chemical vapor deposition. *J. Appl. Phys.* **113**, 203501 (2013).

[29] Altshuler, B. L. and Aronov, A. G. Electron-Electron Interactions in Disordered Conductors (Elsevier, Amsterdam, 1985).

[30] Piquemal, F. *et al.* Report on a joint bipm-euromet project for the fabrication of qhe samples by the lep. *IEEE Trans. Instrum. Meas.* **42,** 264 (1993).

[31] Novoselov, K. S. *et al.* Room-temperature quantum Hall effect in graphene. *Science* **315** 1379–79 (2007).

[32] Janssen, T. J. B. M. *et al.* Anomalously strong pinning of the filling factor $\nu = 2$ in epitaxial graphene. *Phys. Rev. B* **83,** 233402–6 (2011).

[33] Lafont, F. *et al.* Quantum Hall resistance standard based on graphene grown by chemical vapor deposition on silicon carbide. *Nat. Comms*. **6**, 6806 (2015).

[34] Bloch, F., Cugat, O., Meunier, G, Toussaint J.C. Innovating approaches to the generation of intense magnetic fields : design and optimization of a 4 tesla permanent magnet flux source. *IEEE Transactions on Magnetics* **34** (5 PART 1), 2465-2468 (1998).

[35] Radebaugh, R. Cryocoolers: the state of the art and recent developments. *J. Phys.: Condens. Matter* **21**, 164219 (2009).

[36] Allan, D.W. Should the classical variance be used as a basic measure in standards metrology? *IEEE Trans. Instrum. Meas.* **36,** 646-654 (1987).

[37] Kibble, B.P. A measurement of the gyromagnetic ratio of the proton by the strong field method. *Atomic Masses and Fundamental Constants* vol **5**, editors (J.H. Sanders and A.H. Wapstra (New York 1976) 545-51.

[38] Hartland, A., Jones, K., Williams, J.M., Gallagher, B.L., Galloway, T. Direct comparison of the quantized Hall resistance in gallium arsenide and silicon. *Phys. Rev. Lett*. **66,** 969 (1991).

[39] Novoselov, K. S. *et al.* A roadmap for graphene. *Nature* **490,** 192–200 (2012).

[40] Lara-Avila, S. *et al.* Non-volatile photochemical gating of an epitaxial graphene/polymer heterostructure. *Adv. Mater.* **23**, 878 (2011).

[41] Jabakhanji *et al*, Tuning the transport properties of graphene films grown by CVD on SiC(0001): Effect of in situ hydrogenation and annealing. Phys. Rev. B **89**, 085422 (2014).

[42] Kedchedzi, K., Fal'ko, V. I., McCann, E., and Altshuler, B. L. Influence of Trigonal Warping on Interference Effects in Bilayer Graphene. *Phys. Rev. Lett.* **98**, 176806 (2007).

[43] Schopfer, F., Poirier, W. Testing universality of the quantum Hall effect by means of the Wheatstone bridge. *J. of Appl. Phys.* **102,** 054903 (2007).

[44] Jeckelmann, B., Jeanneret, B. The quantum Hall effect as an electrical resistance standard. *Rep. Prog. Phys.* **64,** 1603 (2001).

Supplementary materials for

# Graphene surpasses GaAs/AlGaAs

# for the application of the quantum Hall effect in metrology

R. Ribeiro-Palau, F. Lafont, J. Brun-Picard, D. Kazazis, A. Michon, F. Cheynis, O. Couturaud, C. Consejo, B. Jouault, W. Poirier, F. Schopfer*

*Correspondence to: felicien.schopfer@lne.fr

## Materials and Methods

### 1. *Vocabulary of Metrology*

Measurement uncertainty: non-negative parameter characterizing the dispersion of the quantity values being attributed to a quantity intended to be measured, based on the information used.
Measurement accuracy: closeness of agreement between a measured quantity value and a true quantity value of a quantity intended to be measured.
Measurement precision: closeness of agreement between indications or measured quantity values obtained by replicate measurements on the same or similar objects under specified conditions.
$u_C$ is the combined standard uncertainty: standard uncertainty of the result of a measurement when that result is obtained from the values of a number of other quantities, equal to the positive square root of a sum of terms, the terms being the variances or covariances of these other quantities weighted according to how the measurement result varies with changes in these quantities.
In the main text, these quantities are expressed as a relative value and all uncertainties are standard uncertainties given within one standard deviation (1s.d.).
For more information, see the International Vocabulary of Metrology http://jcgm.bipm.org/vim/en/index.html

### 2. *Graphene growth*

Graphene was grown by propane/hydrogen chemical vapor deposition (CVD) on the Si-face of a semi-insulating 0.16° off-axis 6H-SiC substrate from TankeBlue. We used a horizontal hot-wall CVD reactor with a hydrogen/argon mixture (44 % of hydrogen) as the carrier gas at a pressure of 80 kPa during the whole process. The graphene growth was obtained by adding a propane flow (0.028 %) for 10 min at a growth temperature of 1550°C *(26).*

### 3. *Device fabrication*

The graphene sample was annealed in vacuum (about 10 mPa pressure) for 1 min at 500°C (ramp of 500 s). The sample was left to cool down to below 100°C in vacuum over a few minutes. Subsequently, it was covered with polymethylmethacrylate (PMMA) for protection. The Hall bar was patterned in the form that can be seen in Fig. 2a using electron-beam lithography and PMMA resist. The Hall bars were defined using oxygen reactive ion etching (RIE). Ohmic contacts to the graphene layer were formed by depositing a Pd/Au (60/20 nm) bilayer in an electron beam deposition system, using an ultra-thin Ti layer for adhesion. Thicker Ti/Au (20/200 nm) bonding pads were formed in a subsequent step, where a RIE etch was performed prior to metal deposition for better adhesion of the metal pads to the SiC substrate.
Finally, the sample was covered by 300 nm of poly(methylmethacrylate-co-methacrylate acid) copolymer (MMA (8.5) MAA EL10 from Microchem) and 300 nm of poly(methylstyrene-co-chloromethylacrylate) (ZEP520A from Zeon Chemicals) resist *(40)*. The ZEP520A resist is known to reduce the electron density under UV illumination. Nonetheless, no illumination was done in our case.

## 4. ***Graphene characterization***

The graphene layer grown by CVD on SiC was characterized by atomic force microscopy in tapping mode. The topology (Fig. S1, left) presents SiC steps defining terraces with typical width of 300 nm. The small bright points located near the SiC steps are attributed to external contamination. For some terraces, we can observe on the phase image (Fig. S1, center) a bright contrast attributed to second graphene layer coverage. Low energy electron diffraction and microscopy (LEED and LEEM) were performed on a graphene sample grown simultaneously with the sample used to fabricate the main device presented in this article, but using n-type doped substrate (instead of semi-insulating substrate for the main sample) in order to ease the analysis. LEED (not shown) establishes the presence of a $(6\sqrt{3}\times6\sqrt{3})$-R30° reconstructed interface layer beneath the graphene layer. The LEEM image (Fig S1, right) presents bright and dark contrasts. Reflectivity measurements allow to attribute bright contrast to single layer coverage and dark contrast to bilayer, making LEEM observation consistent with AFM phase analysis. From AFM and LEEM, we can deduce that small micrometric bilayer inclusions, scattered over the surface, cover roughly 15 % of the sample surface.

Additionnal structural characterizations performed in samples growth by the same technique have been published in ref. *(41).*

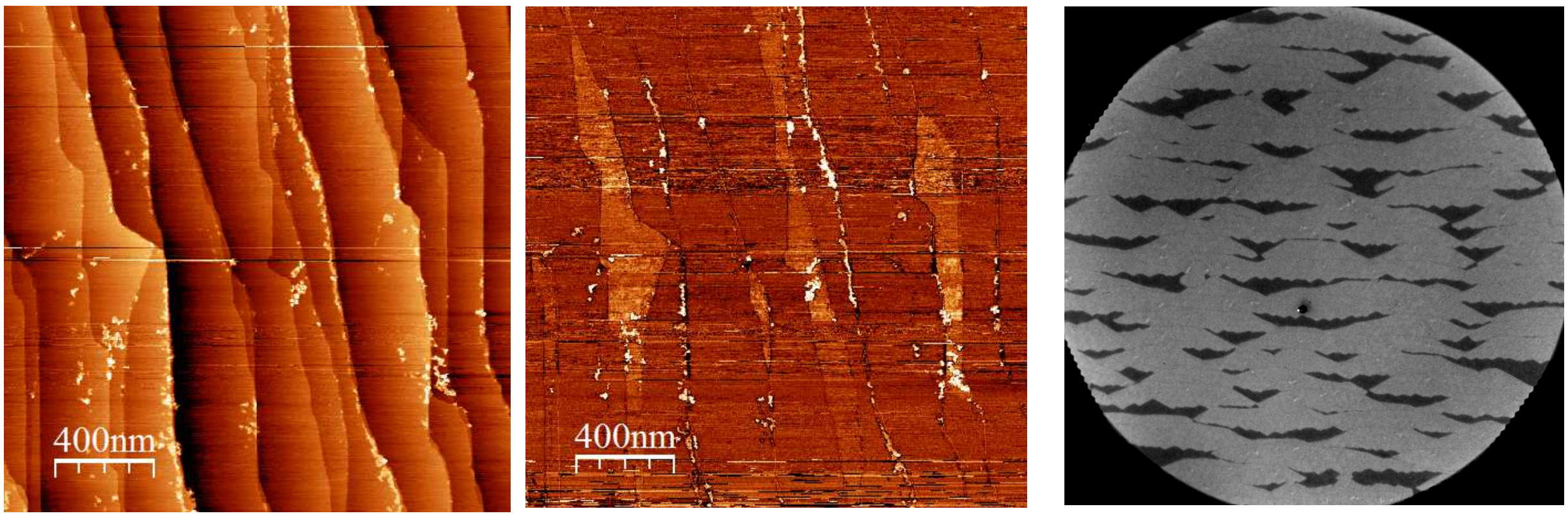

**Figure S1.** Images of the graphene layer grown by CVD on SiC by atomic force microscopy (Height color scale from black to white is 2.5 nm), topology (left), phase (center) and by low energy electron microscopy (diameter of the disk is 10 µm) (right).

## 5. ***Detailled description of the electronic transport measurements carried out including the experimental techniques***

### 5.1. Charge carrier density and mobility at low magnetic field

The charge carrier density at low magnetic field has been extracted by two distinct usual ways: from the magnetic field dependence of the four-terminal Hall (transverse) resistance in the classical Hall effect regime, on one hand, and from the magnetic field dependence of the Shubnikov de Haas oscillations (SdH) of the four-terminal longitudinal resistance (Fig. 2b). The Hall coefficient gives an electron density of $n_s$~$1.8\times10^{11}$ cm$^{-2}$, at low temperature (1.3 K). Though more difficult with our data, analysis of the SdH oscillations confirm this density value. The carrier density is remarkably homogeneous at low magnetic field: typical discrepancies by $1.5\times10^{10}$ cm$^{-2}$ are observed by comparing the measurements carried out with different voltage terminal-pairs.
Knowing the carrier density, the carrier mobility is extracted from the Drude contribution to $R_{xx}$ at $B\approx0$ (obtained by subtraction of the well-known quantum corrections to electronic transport, namely the weak-localization peak combined with the contribution of electron-electron interaction. We found $\mu$~9 400 cm$^2$V$^{-1}$s$^{-1}$. Another indication of this rather high charge mobility is the onset of the SdH oscillations at 1 T. Actually, the longitudinal resistance is expected to start to oscillate with the magnetic field as soon as the cyclotron orbits becomes smaller than the electron mean free path (i.e. the Landau levels becomes well separated), leading to $\mu B>1$.

### 5.2. Weak-localization correction measurements

Magnetoresistance at low magnetic field was analysed with the appropriate theory of weak-localization *(42)*. Below 7 K, we observe a saturation of the extracted phase coherence length $L_\phi$ at about 300 nm. This length can be compared to the typical width of the SiC terraces.
The characteristic length of intervalley scattering $L_i$, which is also deduced from this analysis, is temperature independent and equal to about 140 nm. Much below $L_\phi$ value, $L_i$ value confirms the presence of sharp short-range disorder, what could be explained by the presence of bilayer inclusions. The characteristic length of intravalley scattering process $L_*$, also determining the magnetoresistance associated with the weak-localization correction, is of the same order of magnitude as the transport length equal to $l_{tr}$=45 nm.

### 5.3. Resistance of the electrical contacts to the graphene device

The resistance of each of the eight metallic Ti/Pd/Au contacts to the graphene was measured on the Hall resistance plateau at $h/(2e^2)$ by a three-terminal technique. This technique consists in using the contact to be probed both to inject the measurement current and to probe the potential, a second contact to drain the measurement current, and a third contact to probe a second potential. In the QHE regime, assuming the contacts used to probe the potential are chosen along the same edge state/equipotential, the resistance measured equals $R_L+R_c+r_{xx}$, $R_L$ is the known resistance of the line connecting the room-temperature connector at the top of the cryoprobe to the contact on the sample, $R_c$ is the resistance of the contact itself, and $r_{xx}$ is the longitudinal resistance between the two voltage probes which is very close to zero in the QHE regime *(4)*.
The resistance of each of contacts to the graphene was found to be $R_c\leq1.1\ \Omega$, as it is show in Fig. 2a. It does not depend on the magnetic field over the tested range (3.5 T – 14 T). For the largest source and drain contacts, the resistance is below 100 m$\Omega$ and it does not depend on the current, at least, up to 550 μA. The resistance of the smaller voltage contacts was measured with currents up to 20 μA and showed no current dependence over this range. Such a contact quality is an asset for the measurement of a quantized Hall resistance with the best precision.

### 5.4. Longitudinal and Hall resistance measurements in the QHE regime

For all the reported four-terminal resistance measurements of the graphene device, the current circulates between the source and the drain at the ends of the Hall bar (Fig. 2a). The longitudinal resistance $R_{xx}$, except for the measurements reported in Figure 2b, corresponds to the mean of the four values using terminal-pairs (A, B), (B, C), (E, F), (F, G). In Figure 2b, $R_{xx}$ is deduced from measurements of the voltage between (F, G), as indicated on the picture of the sample. The longitudinal resistances indicated in the main text are normalized to a square geometry.

#### *5.4.1. Precise Hall resistance measurements:*

The accuracy of the quantized Hall resistance in the graphene device, $R_{H\text{-}G}$, was demonstrated through indirect comparisons of $R_{H\text{-}G}$ with $R_{H\text{-}GaAs}$ delivered by GaAs-QHRS, *via* an intermediate 100-Ω room-temperature resistor and using a cryogenic current comparator (CCC) bridge *(4)*.

The CCC is a perfect transformer which can measure a current ratio in terms of the winding number of turns ratio with a relative uncertainty as low as a few $10^{-11}$. Its accuracy relies on a flux density conservation property of the superconductive toroidal shield (Meissner effect), in which superconducting windings are embedded and on a flux detector based on a DC superconducting quantum interference device (SQUID).

The principle of the resistance bridge (Fig. S1) is to compare the currents flowing in the two resistors to be compared while the voltage drops at the terminals of the two resistors are balanced *(4)*.
It is based on:
- two servo-controlled current sources feeding the two resistors;
- a cryogenic current comparator (CCC) based on a SQUID used to servo-control the ratio of the two currents circulating through two of its windings;
- a calibrated current divider used to further adjust the ratio of the two currents;
- a null detector to check the voltage balance at the terminals of the two resistors.

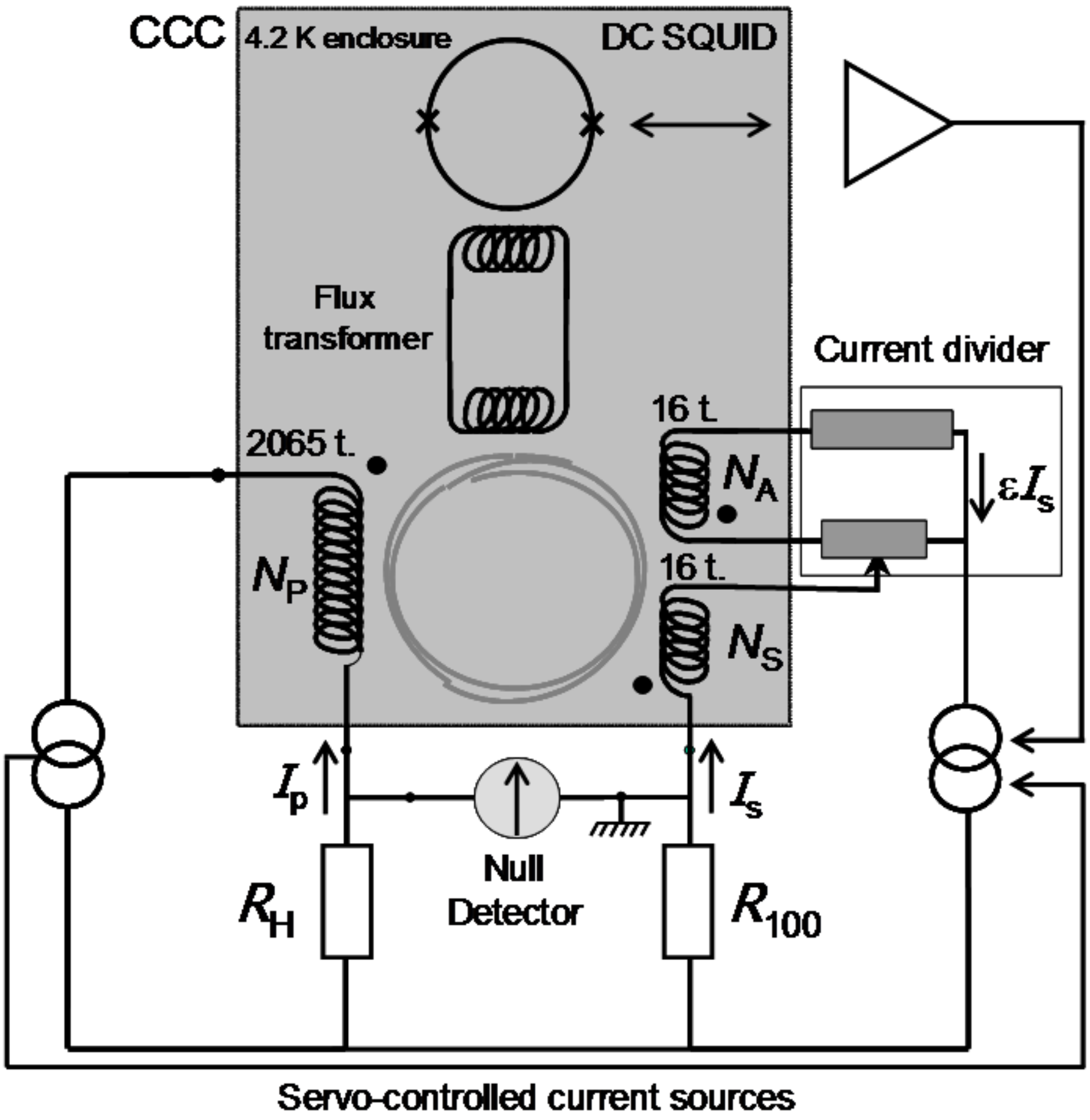

**Figure S2.** Schematic of the resistance bridge circuit based on a cryogenic current comparator.

The ratio of the currents is coarsely pre-adjusted so that the voltage at the input of the null detector is only a few $10^{-6}$ of the voltage drop at the resistor terminals. $I_P$ and $I_S$ currents, which feed the resistors, also circulate across windings of the CCC of number of turns $N_P$ and $N_S$ respectively. To compare a Hall resistance $R_H$ quantized at $h/(2e^2)$ and a 100-Ω resistor of resistance $R_{100}$, we connect them in series with a $N_P$=2065 turns winding and a $N_S$=16 turns winding, respectively. In this case, the ratios $N_P/N_S$ and $R_P/R_S$ are nearly equal within $1\times10^{-5}$. This choice ensures that the flux measured by the CCC SQUID is low. An auxiliary winding of number of turns $N_A$ is supplied with a fraction ε of the current $I_S$ by using the calibrated current divider. In external feedback operating mode of the SQUID, $I_S$ is regulated so that the flux measured by the SQUID is maintained at zero and $\varepsilon_0$ ensures that the null detector measures zero voltage. This results in:

$$R_H\, I_P = R_{100}\, I_S$$

$$N_P I_P = N_S I_S (1+\varepsilon_0 N_A/N_S)$$

It comes:

$$\gamma^{-1} = R_{100}/R_H = N_S/N_P\,(1+\varepsilon_0 N_A/N_S)$$

Finally, the resistance bridge enables the determination of the ratio γ of two resistances, in terms of ratios of winding numbers of turns and a fraction $\varepsilon_0$ of the calibrated current divider.

The determination of $\varepsilon_0$ is made from two calibrated values of $\varepsilon+$, $\varepsilon-$ , which respectively discards from $\varepsilon_0$ by about $2.5\times10^{-7}$, and respectively gives rise to voltage unbalance, $\Delta V+$, $\Delta V-$ of opposite signs, and precisely measured with the null detector. $\varepsilon_0$ is then deduced from the linear interpolation function between $\Delta V+(\varepsilon+)$ and $\Delta V-(\varepsilon-)$, denoted $\Delta V(\varepsilon)$, as $\Delta V(\varepsilon_0)=0$.

$\Delta V\pm$, is measured in quasi-DC mode, with the current direction periodically reversed (a delay (dead time) is kept after each current reversal and before measurement of the voltage unbalance in each current direction $I+$ or $I-$). The current reversal allows to eliminate electrical offsets, drifts, and possible other non-ohmic parasitic signals.

Owing to a flux detector based on a Quantum Design DC superconducting quantum interference device (SQUID), the current white noise of the CCC is 80 pA.turn/Hz$^{1/2}$. For all measurements reported, the resistance bridge is equipped with EMN11 nanovoltmeter as a null detector to measure the voltage unbalance at the terminals of the two resistances to be compared. The voltage white noise of this null detector is 7 nV/Hz$^{1/2}$. It is the dominant contribution to the final uncertainty of a measurement of a resistance ratio $\gamma$. The typical standard uncertainty, evaluated from series of repeated observations (type A evaluation), of a measurement of $\gamma\,(=R_H/R_{100})$ is below $0.3\times10^{-9}$, for a 30 minutes measurement, and with a measurement current of 50 μA in $R_H$.

The room-temperature resistor used in the indirect comparison of $R_{H\text{-}G}$ and $R_{H\text{-}GaAs}$ is a 100-Ω resistor (TEGAM) of resistance $R_{100}$. It is a well-characterized material standard, in a shielded and temperature-controlled enclosure adjusted at 21.6°C, where the minimum of the quadratic temperature dependence of its resistance occurs (lower than $5\times10^{-9}$/K). Consequently, even the temperature variation of 7 mK observed when the measurement current reaches its maximum used value of 7.74 mA (corresponding to 60 μA in the QHRS on the so-called $h/(2e^2)$ plateau), leads to negligible variations of the resistance of the standard, with regards to the measurement uncertainties reported. Anyway, the measurement protocol of the indirect comparison between $R_{H\text{-}G}$ and $R_{H\text{-}GaAs}$, as described below, would have cancelled any possible reproducible resistance change induced by heating with measurement current. The stability in time of the material standard resistance has been fully characterized. At long-time scale (3 months), the mean value of the resistance drift is below $+7\times10^{-11}$ per day, in relative value. The typical stochastic instability daily observed is covered by the combined standard uncertainty of the measurements ($<5\times10^{-10}$).

The GaAs-QHRS (LEP 514) *(22)* used in the comparison is made from a two-dimensional electron gas, obtained in a GaAs/AlGaAs semiconductor heterostructure, with density $5.2\times10^{11}$ cm$^{-2}$ and mobility 283000 cm$^2$V$^{-1}$s$^{-1}$, patterned into a $w$=400 μm wide Hall bar. It was fully characterized according to the technical guidelines for quantum Hall resistance metrology *(11)*. Notably when placed at 10.8 T and 1.3 K, at $\nu=2$, the resistance of each of the eight Au/Ge/Ni contacts is below 200 mΩ, $R_{xx}$ stays below 10 μΩ up to high measurements currents of 60 μA, Fig. S3. Considering a typical coupling factor of 0.1 between $R_H$ and $R_{xx}$, as observed in previous experiments with these LEP 514 samples at $\nu=2$, within the experimental realization uncertainty *(30)*, this low value predicts that the Hall resistance in the GaAs-QHRS, $R_{H\text{-}GaAs}$, agrees with the most expected quantized value $h/(2e^2)$ with a relative accuracy better than $8\times10^{-11}$. In the experiments, the accuracy of the quantized Hall resistance in the G-QHRS is assessed by comparison with $R_{H\text{-}GaAs}$.

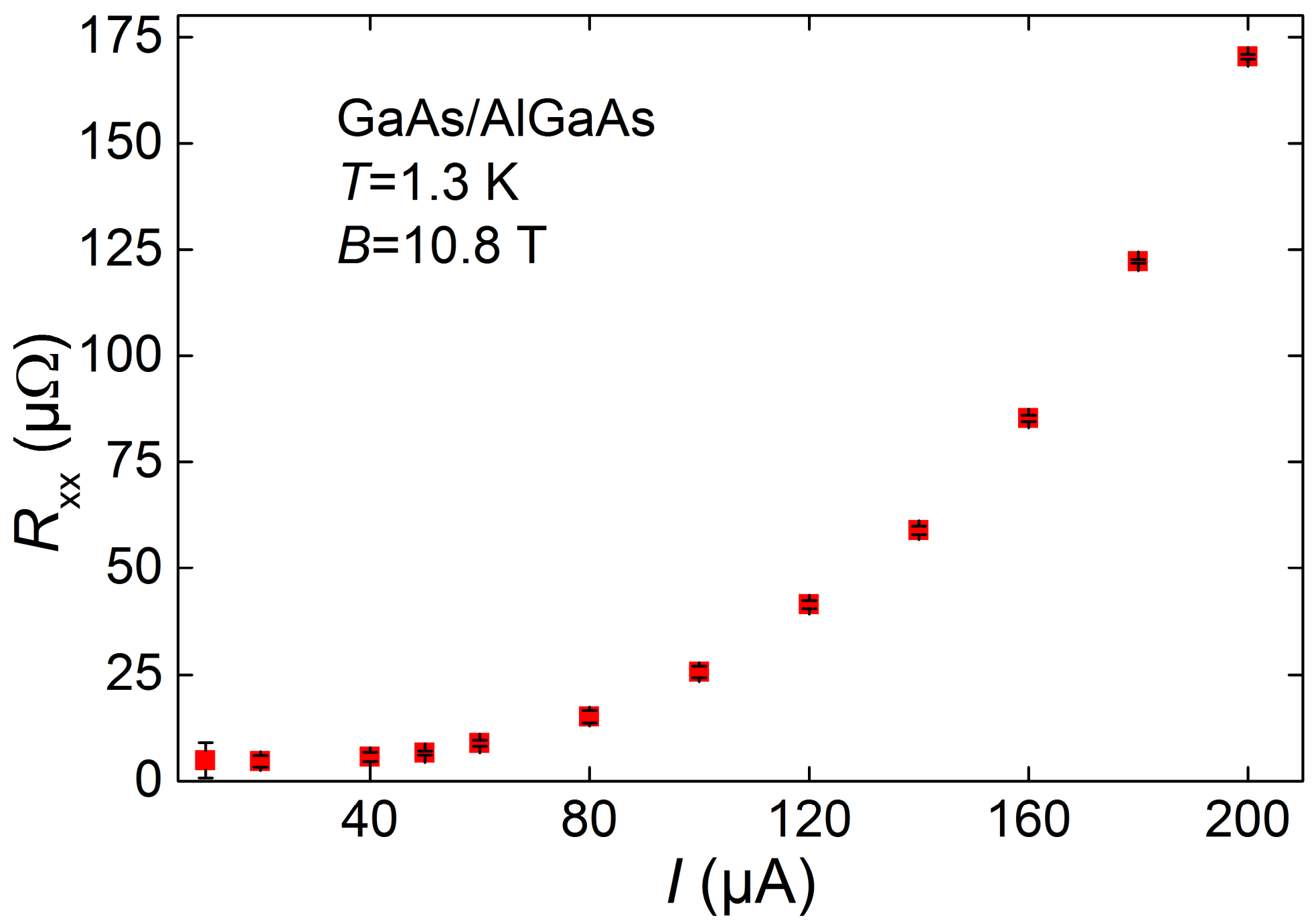

**Figure S3.** Mean value of $R_{xx}$ (per square) as a function of the measurement current $I$ for GaAs-QHRS LEP 514 used in the experiment at 10.8 T, 1.3 K..

In principle, the indirect comparison of $R_{H\text{-}G}$ and $R_{H\text{-}GaAs}$ *via* $R_{100}$, which allows a determination of $R_{H\text{-}G}$ in terms of $R_{H\text{-}GaAs}$, results from two distinct resistance comparisons: the first one between $R_{H\text{-}GaAs}$ and $R_{100}$ and the second one between $R_{H\text{-}G}$ and $R_{100}$. The resistance bridge measures $\gamma_{-GaAs}=R_{H\text{-}GaAs}/R_{100}$ in the first experiment and $\gamma_{-G}= R_{H\text{-}G}/R_{100}$ in the second. Assuming that $R_{100}$ is invariant in the two measurements, it comes $R_{H\text{-}G}\text{-}R_{H\text{-}GaAs}= \Delta R_{H\text{-}G} = R_{100}(\gamma_{-G} - \gamma_{-GaAs})$. In order to check the invariance of $R_{100}$ we have systematically performed measurements of $\gamma_{-GaAs}$ before and after each measurement of $\gamma_{-G}$. Denoted $\gamma_{-GaAs\text{-}1}$ and $\gamma_{-GaAs\text{-}2}$, they were separated by less than 6 hours. We rejected the determinations $\Delta R_{H\text{-}G}$ based on $\gamma_{-GaAs\text{-}1}$ and $\gamma_{-GaAs\text{-}2}$ significantly disagreeing to more than the combined standard uncertainty. Nevertheless, even for the selected determinations of $\Delta R_{H\text{-}G}$, we have systematically considered the mean of $\gamma_{-GaAs\text{-}1}$ and $\gamma_{-GaAs\text{-}2}$, in order to minimize the residual impact of $R_{100}$ stochastic instability and of its drift in $\Delta R_{H\text{-}G}$ measurement. Note that the expected drift of $R_{100}$ over 6 h is negligible $<2\times10^{-11}$, and the $R_{100}$ stochastic instability contributes to random uncorrelated noise in repeated measurements of $\Delta R_{H\text{-}G}$. Considering two measurements $\gamma_{-GaAs\text{-}1}$ and $\gamma_{-GaAs\text{-}2}$ has also the advantage to reduce the final combined standard uncertainty of $\Delta R_{H\text{-}G}$. Each determination of $R_{H\text{-}G}$ in terms of $R_{H\text{-}GaAs}$ is thus deduced from three measurements: $\gamma_{-GaAs\text{-}1}$, $\gamma_{-G}$ and $\gamma_{-GaAs\text{-}2}$. It comes:

$$R_{H\text{-}G}\text{-}R_{H\text{-}GaAs}= R_{100}[\ \gamma_{-G} - (\gamma_{-GaAs\text{-}1}+\gamma_{-GaAs\text{-}2})/2]$$

and

$$(\Delta R_H/R_H)_{-G}=(R_{H\text{-}G}\text{-}R_{H\text{-}GaAs})/R_{H\text{-}GaAs} \approx 100/12906\times[\ \gamma_{-G} - (\gamma_{-GaAs\text{-}1}+\gamma_{-GaAs\text{-}2})/2]$$

(Here, "≈" refers to an approximation to within about $10^{-14}$.)

In the experiments, the GaAs-QHRS and G-QHRS are placed in the same cryomagnetic setup, which combines a $^4$He variable temperature insert and a 14 T superconducting magnet. The cables of the cryoprobe used for the transport measurements of both graphene and GaAs/AlGaAs are of the same type with the same properties and behaviour. Each of the cables is PTFE insulated with an external grounded shielding. The insulating resistance to ground of each cable is higher than $10^{14}$ Ω. Independently of its measurement conditions, $R_{H\text{-}G}$ is compared to $R_{H\text{-}GaAs}$ obtained in the best quantization conditions, *i.e.* 1.3 K and ±10.8 T.

In addition, the two comparisons of $R_{H\text{-}GaAs}$ and $R_{100}$ ($\gamma_{-GaAs\text{-}1}$, and $\gamma_{-GaAs\text{-}2}$), on one hand, and the comparison of $R_{H\text{-}G}$ and $R_{100}$ ($\gamma_{-G}$), on the other hand, combined to deduce a value of $(\Delta R_H/R_H)_{\text{-}G}$ are performed in exactly the same measurement configurations: same current, same cables between the bridge and the top of the cryostat, same configuration of the bridge (including same CCC winding ratio, same position of the current divider used to adjust the current in the resistances, same pre-adjustment of the current ratio in the resistances), same measurement protocol (including time constants and delays). Only the cables of the cryoprobe used for the measurements of $\gamma_{-GaAs\text{-}i}$ and $\gamma_{-G}$ are different. Nevertheless, due to their very close high insulating resistance to ground, and the configuration of the bridge, where the insulating resistance to ground is put parallel to the 100-Ω resistor, their contribution to the measurement uncertainty of $(\Delta R_H/R_H)_{\text{-}G}=(R_{H\text{-}G}\text{-}R_{H\text{-}GaAs})/R_{H\text{-}GaAs}$ is lower than $1\times10^{-11}$. As a consequence, this measurement procedure perfectly subtracts all the possible significant systematic error of each resistance ratio $\gamma$ measurement, in a given $(\Delta R_H/R_H)_{\text{-}G}$ determination. The measurement uncertainty finally only comes from random noise. The residual errors introduced by the small irreproducibilities of the bridge adjustment, from one $\gamma_{-GaAs\text{-}i}$ or $\gamma_{-G}$ measurement to another similar $\gamma_{-GaAs\text{-}i}$ or $\gamma_{-G}$, only negligibly contribute to random noise in the case of the most precise determinations of $(\Delta R_H/R_H)_{\text{-}G}$ (Fig. 6a), because of the combination of a large number of measurements.

The measurement protocol includes alternating the forward and reverse DC current directions (every 35 s), so as to eliminate electrical offsets, drifts, and possible other non-ohmic parasitic signals. A measurement of $\gamma\,(=R_H/R_{100})$, indifferently with G-QHRS or GaAs-QHRS, is typically 30 min to 40 min long (including the measurements of $\Delta V+(\varepsilon+)$ and $\Delta V\text{-}(\varepsilon-)$). Exactly the same measurement protocol, including the time sequence, is used for the determinations of $\gamma_{-G}$ and $\gamma_{-GaAs}$.

Figure S4 shows a typical measurement sequence of $\Delta V+(\varepsilon+)$, for $\varepsilon+=-12.15\times10^{-6}$, in the context of the determination of $\gamma_{-G}$, with 50 μA circulating in the graphene device placed at 1.3 K and 5 T, and using the central transversal terminal-pair (B, F). A voltage drop of about 0.645 V develops at the terminal-pair used, in these conditions ($R_{H\text{-}G}$ quantized on $h/(2e^2)$ plateau). $\Delta V+$ is deduced from the series of values $\Delta V+_i$ obtained after two current reversals. The subsequence considered for the determination of one elementary value $\Delta V+_i$, is delimited by a dashed black line on Fig S4. $\Delta V+_i = [<\Delta V+(I+)>-<\Delta V+(I\text{-})>]/2$, where $<\Delta V+(I\pm)>$ is the average of all the $\Delta V+(I\pm)$ measurements over the time interval of the subsequence. The subsequence is design to consider measurements of $\Delta V+(I+)$ and $\Delta V+(I\text{-})$ over equal time interval, and half of the $\Delta V+(I+)$ measurements performed before and after $\Delta V+(I\text{-})$ measurements, so that any drift with time is efficiently subtracted.
Note that the data presented on Fig S4 is a visualization of the null detector measurement on a plotting table connected to the analog output. The thickness of the line is related to the ink pen quality rather than the measurement noise. The determination of $\Delta V+$ is not done from the data presented, but directly from the null detector measurements recorded with a multimeter.

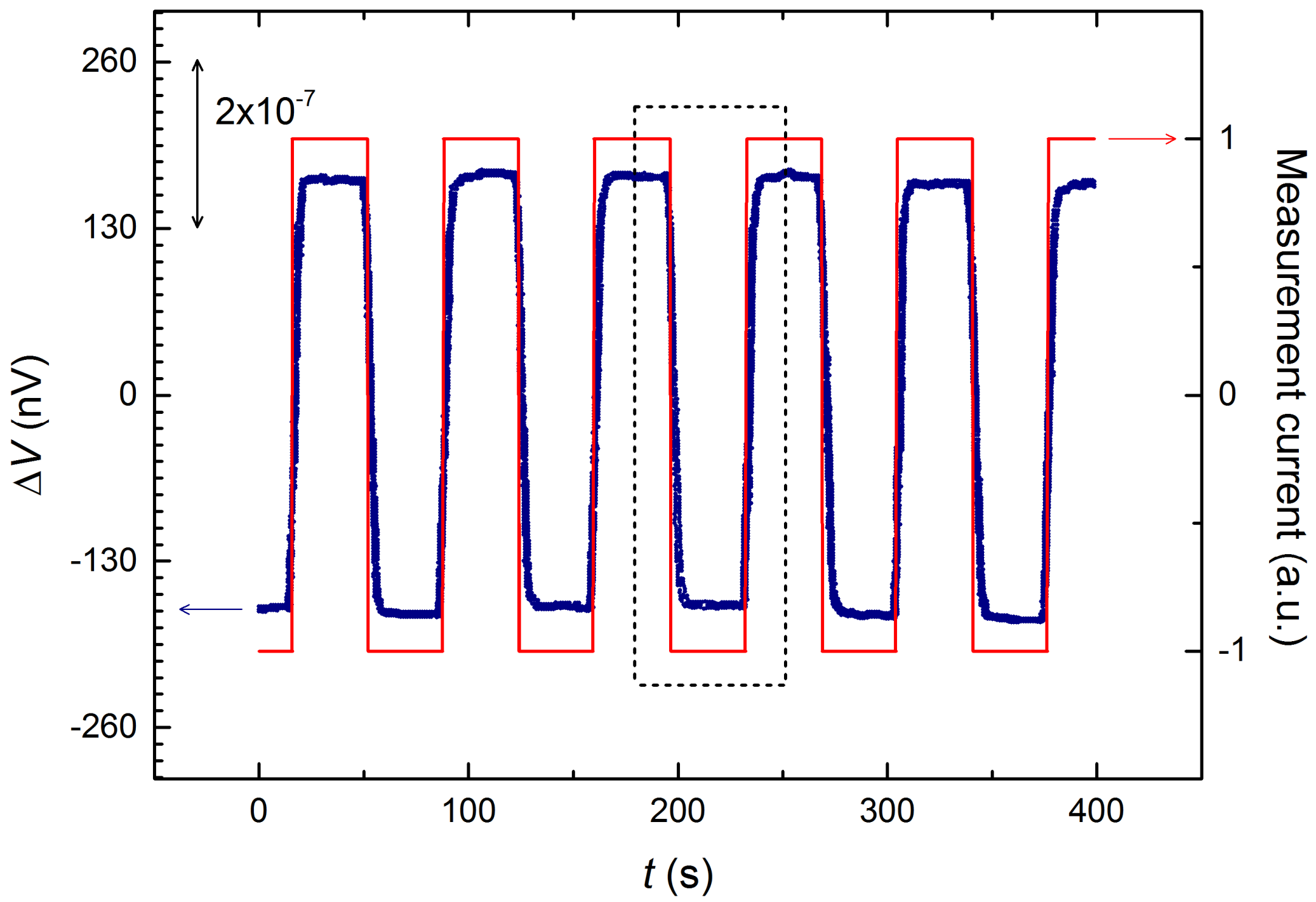

**Figure S4.** Typical measurement sequence of $\Delta V$+(ε+), for ε+=-12.15×$10^{-6}$, with EMN11 as null detector, in the context of the determination of $\gamma_{-G}$, with a measurement current of 50 μA circulating in the graphene device placed at 1.3 K and 5 T, and using the central transversal terminal-pair (B, F). A voltage drop of about 0.645 V develops at the transversal terminal-pair used, so that 130 nV unbalance voltage corresponds to a 2×$10^{-7}$ fraction of the resistance ratio. $\Delta V$+ is deduced from the series of values $\Delta V+_i$ obtained after two current alternations. The subsequence considered for the determination of one elementary value $\Delta V+_i$, is delimited by a dashed black line. $\Delta V$+ is schematically deduced from [$\Delta V$+($I$+)-$\Delta V$+($I$-)]/2. $\Delta V+_i$ = [<$\Delta V$+($I$+)>-<$\Delta V$+($I$-)>]/2, where <$\Delta V$+($I$±)> is the average of the $\Delta V$+($I$±) measurements over the time interval of the subsequence. Current reversal occurs every 35 s.

We have checked that the noise in a measurement of resistance ratio γ ($\gamma_{-GaAs\text{-}i}$ or $\gamma_{-G}$) is predominantly white (uncorrelated random), as proven by calculation of the Allan deviation *(36)*. As shown on Fig. S5 for a measurement of $\gamma_{-G}$ with G-QHRS at 5 T, 1.3 K, using 50 μA, the Allan deviation, calculated from the elementary measurements $\Delta V\pm_i$, depends on the measurement time $t$ following $t^{-1/2}$, as expected in the case of white noise. Hence there is a good agreement between Allan deviation calculations and calculations of the experimental standard deviation of the mean. Note that for 38 min (60 min) measurement of the graphene device, we achieve a relative standard uncertainty of 0.2×$10^{-9}$ (0.15×$10^{-9}$), at the level of the best capability of the bridge. Similar Allan variance and relative standard uncertainty have been obtained for $\gamma_{-GaAs}$.

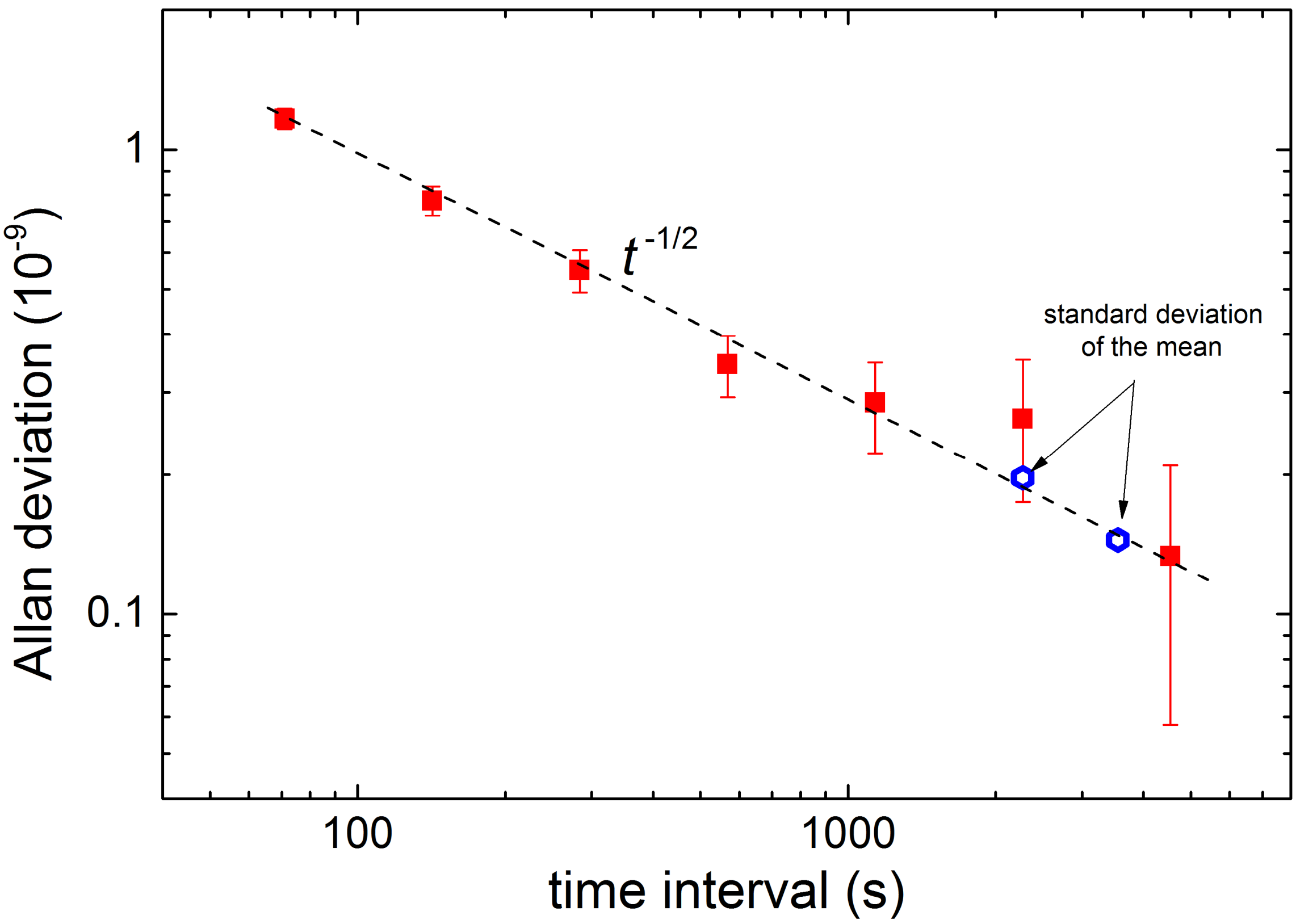

**Figure S5.** Allan deviation (red points) and experimental standard deviation of the mean (blue points) for a measurement of $\gamma_{\text{-G}}=R_{\text{H-G}}R_{100}$, as a function of the measurement time. The $t^{-1/2}$ dependence (dashed black line) indicates white noise in the measurement. G-QHRS is measured, in this case at 5 T, 1.3 K, using 50 µA.

### *5.4.2. Precise longitudinal resistance measurements:*

Precision measurements of $R_{xx}$ both in GaAs/AlGaAs and graphene were performed using a SQUID and the CCC as an amplifier. As for $R_H$ measurements, the data are collected in alternating the DC current directions.

A 2065 turns winding of the CCC is connected to the two longitudinal voltage terminals. The longitudinal voltage gives rise to the circulation of a current $i_{xx}$ in the winding of the CCC with a SQUID operating in internal feedback mode. The output of the SQUID electronics is measured with an Agilent 3458A multimeter. The current noise resolution is about 40 fA.Hz$^{1/2}$ which results in a voltage noise resolution of about 0.5 nV Hz$^{1/2}$. The longitudinal resistance $R_{xx}$ is then given by $R_{xx} = (i_{xx}/I)R_H$, with $I$ the measurement current, since the two-terminal impedance seen by the winding is very close to $R_H$ on the quantized Hall plateau.

## *6. Further description of the Hall quantization state in G-QHRS*

### 6.1. Coupling between $R_H$ and $R_{xx}$ and homogeneity properties.

As shown on Fig. 3a and b, we observe an obvious correlation between $(\Delta R_H/R_H)_{\text{-G}}(B)$, measured with the central transversal terminal-pair (B,F), and $R_{xx}(B)$. This is more clearly confirmed in case of significant deviation of $R_{\text{H-G}}$ from $R_{\text{H-GaAs}}$ and large values of $R_{xx}$ as observed at higher current, at 3.5 T and 4 T. It is possible to describe this correlation, at a given $B$, by a linear coupling between $\Delta R_{\text{H-G}}(B)$, and $R_{xx}(B)$, with a constant coefficient of 0.2 to 0.4, independent of $B$, over the tested range 3.5 T – 14 T, on the $h/(2e^2)$ plateau, for a given direction of the magnetic field. It manifests a parasitic contribution of $R_{xx}$ to $R_{H\text{-}G}$ measurement. Such a coupling, already observed in metrological samples *(44)*, can be explained by two dominant mechanisms *(43)*:

i) A purely geometric effect caused by the finite width $w_p$ of the voltage probe as compared to the Hall bar width $w$, and on the chirality of the currents flowing along the edges (with respect to $B$-direction).
ii) The inhomogeneity of carrier density or the disorder resulting in the distortion of the current flowing.

Consequently the coupling can be described in a simple model by

$$R_{H\text{-}G} = h/(2e^2) + (\alpha_{geo} + \alpha_{inhomo})\, R_{xx}.$$

Because of the chirality of edge currents between edge states, $\alpha_{geo}(\pm B) = -\, w_p/w$ is even with respect to the $B$-direction, *i.e.* independent on the $B$-direction and, $\alpha_{inhomo}(\pm B) = \pm\sin(\beta)$, with $\beta$ the tilt angle of the current flow with respect to the Hall bar orientation, is odd.

For the given geometry of the G-QHRS, $\alpha_{geo} = -0.2$, so that the observed coupling factor of 0.2 to 0.4, can only be explained by considering the additional coupling $\alpha_{inhomo} = +0.4$ to +0.6, in this magnetic field direction. The existence of this coupling term confirms a certain degree of (microscopic) inhomogeneity in the device. Such inhomogeneity could be related to the SiC steps, considering that they are able to drive the current flowing. In this case, the experimental values of $\alpha_{inhomo}$ are compatible with SiC steps aligned following directions forming angles between 25° to 40° with the Hall bar orientation.

Inverting the magnetic field direction reveals a coupling factor equal to -0.8 to -0.6, the sum of $\alpha_{geo} = -0.2$ and $\alpha_{inhomo} = -\,0.6$ to -0.4, which changes its sign. This validates the model.

The observation of a coupling between $\Delta R_{H\text{-}G}(B)$, and $R_{xx}(B)$ following this model is remarkable. It has also been done in ref. *(33)*. The $R_{H\text{-}G}$ does not present a finite slope which crosses the ideal value near the minimum of dissipation (minimum of $R_{xx}$) measured between 5 T - 6 T, as observed in certain cases *(43)*.

Inverting the magnetic field direction is also very instructive about the Hall quantization accuracy, mainly because revealing these detrimental contributions of $R_{xx}$ to $R_H$, which leads to deviation from the quantization. Beyond, combining $R_H$ measurements in both magnetic field directions allows the precise evaluation of the dissipative transport ($r_{xx}$), directly affecting the Hall quantization accuracy at the location where $R_H$ is probed. This local evaluation of $r_{xx}$ is a better criterion of the Hall quantization accuracy than the measurement of the mean value of $R_{xx}$ along the edges of the sample, particularly in case of inhomogeneity. Figure S6a shows, for the central transversal pair (B,F) of G-QHRS:

$$\Delta R_H/R_{H\text{-odd}} = (R_{H\text{-}G}\,(B+) - R_{H\text{-}G}\,(B-))/(2\, R_{H\text{-}GaAs})$$

$$\Delta R_H/R_{H\text{-even}} = (R_{H\text{-}G}\,(B+) + R_{H\text{-}G}\,(B-))/(2\, R_{H\text{-}GaAs})-1$$

$\Delta R_H/R_{H\text{-even}}$ is nothing else but the mean of the deviations of $R_{H\text{-}G}$ from $R_{H\text{-}GaAs}$ measured in the both magnetic field directions. We rewrite

$\Delta R_H/R_{H\text{-odd}} = \alpha_{inhomo}\ r_{xx}/R_{H\text{-GaAs}}$

$\Delta R_H/R_{H\text{-even}} = \alpha_{geo}\ r_{xx}/R_{H\text{-GaAs}}$

These two quantities turn out to be useful for very precise and reliable indicators of the quantization accuracy. The fact that both $\Delta R_H/R_{H\text{-odd}}$ and $\Delta R_H/R_{H\text{-even}}$ are zero within $u_c = 0.4\times10^{-9}$, from 4 T to 8 T, at 1.3 K, for the central pair, as shown on Fig. S6a, demonstrates that $r_{xx}$ is very low (below 9 μΩ) and that it does not affect $R_{H\text{-G}}$ quantization whatever the coupling mechanism (based on geometry or inhomogeneity) within this uncertainty of $0.4\times10^{-9}$. This information is consistent with the measurement of the mean of $R_{xx}$ measured along the whole circumference of the device, as shown on Fig. 3b. This confirms the good spatial homogeneity of the device at large scale.

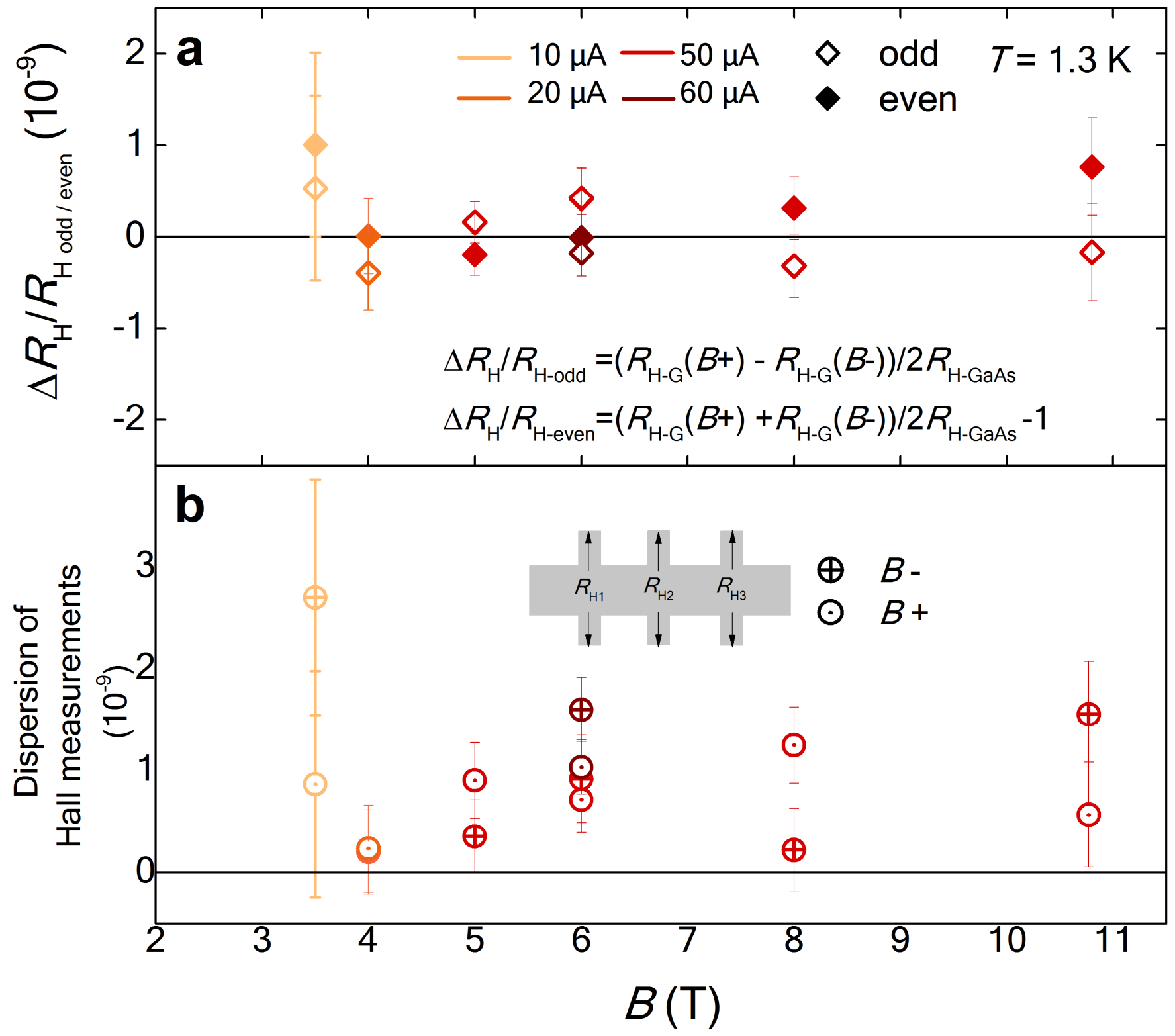

**Figure S6. Robustness of the Hall resistance quantization upon magnetic field direction inversion / Spatial homogeneity of the Hall resistance quantization.** (**a**) $\Delta R_H/R_H$ odd (empty symbols) and even (filled symbols), as defined on the figure, versus magnetic field, for the measurements performed using the central transversal terminal-pair (B, F). $\Delta R_H/R_{H\text{-even}}$ is the mean value of the $(R_{H\text{-G}}\text{-}R_{H\text{-GaAs}})/R_{H\text{-GaAs}}$ measured in the two magnetic field directions. (**b**) Standard deviation of the measurements of $R_{H\text{-G}}$ carried out with the three transversal pairs versus magnetic field, and for both directions. The error bars correspond to the combined standard uncertainty of the standard deviation.

Regarding the spatial homogeneity of the Hall quantization in the device, the dispersion (standard deviation) between $R_{H\text{-G}}$ measurements carried out using the three transversal terminal-pairs of the device (A, G), (B, F), (C, E), each separated by 100 μm, is excellent, below $1\times10^{-9}$ at 4 T and 5 T (for both magnetic field directions), at 1.3 K, and stay below $2\times10^{-9}$ over the wide range 4 T – 10.8 T, at 1.3 K and with currents up to 60 µA (Fig. S6b). The homogeneity of the Hall quantization state at large scale is confirmed. The observation of the small dispersions between these $R_{H\text{-}G}$ measurements also confirms the good robustness of the Hall quantization state in this device, notably against the variation of the measurement probes.

More explicitly, Figure S7 summarizes the quantized Hall resistance measurements performed with G-QHRS using the three transversal terminal-pairs (A, G), (B, F), (C, E), in the magnetic field range 3.5 T - 10.8 T. The upper part shows $(\Delta R_H/R_H)_{\text{-}G}$ for one given direction of the magnetic field. None of the three pairs shows significant deviation of $R_{H\text{-}G}$ from $R_{H\text{-}GaAs}$ within $u_c$, equal to $1\times10^{-9}$, at most, in the conditions of temperature and current indicated. The lower part reports $\Delta R_H/R_{H\text{-}odd}$ and $\Delta R_H/R_{H\text{-}even}$. They are presented as quantization indicators, as explained previously.
The data presented in this figure for the pair (B, F) is the same as presented in the main text in Fig. 3a and Fig. S6a. They are recalled to ease the comparison with the measurements performed with the other pairs.
Remarkably, the measurements of $R_{H\text{-}G}$ using the three transversal terminal-pairs show similar behaviours as a function of the magnetic field, which is correlated with the overall variations of $\Delta R_H/R_{H\text{-}odd}$ and $\Delta R_H/R_{H\text{-}even}$, *i.e.* of the local evaluation of the dissipative transport: an optimum for the Hall quantization accuracy is measured around 5 T – 6 T. In increasing the magnetic field above 6 T, the agreement of the measurements of $R_{H\text{-}G}$ using the three transversal terminal-pairs slightly degrades (while staying below $1.5\times10^{-9}$ up to at least 10.8 T), a certain inhomogeneity manifests itself. The central pair enables the measurement of the most robust quantized Hall resistance accuracy, over the considered magnetic field range. It could be related to the better quality (lower resistance) of the metallic contacts to graphene for this pair.

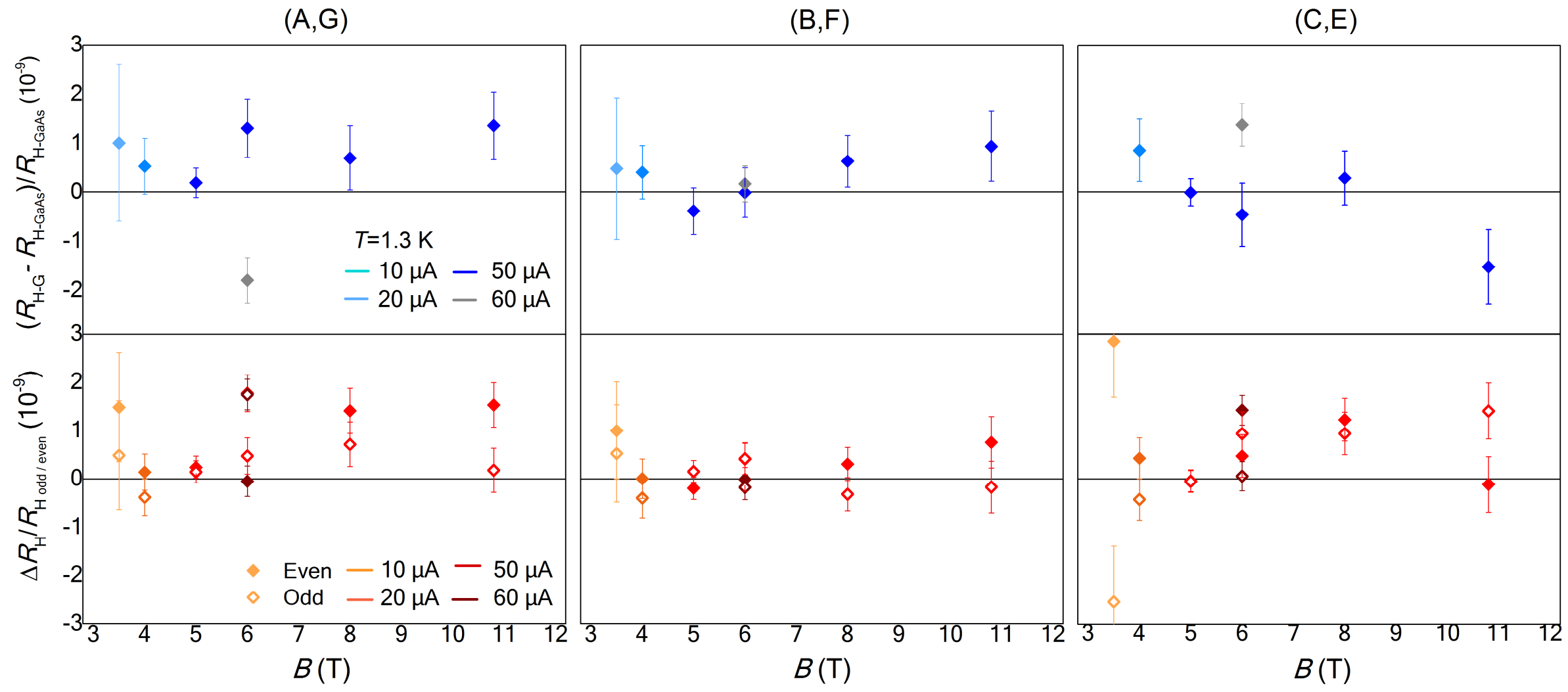

**Figure S7.** Upper part: Relative deviation of $R_{H\text{-}G}$ from $R_{H\text{-}GaAs}$ versus magnetic field, measured using the terminal-pairs (A,G), (B,F), (C,E). Lower part: $\Delta R_H/R_H$ odd (empty symbols) and even (filled symbols) versus magnetic field, for the measurements performed using the terminal-pairs (A,G), (B,F), (C,E).

### 6.2. Additional information about the dependence of $R_{H\text{-}G}$ on current.

The robustness of the Hall quantization accuracy to high measurement currents was directly tested with success by measurements of $R_{H\text{-}G}$ up to 60 µA. The perfect agreement with the value at 50 µA (see data at 6 T, 1.3 K on the Fig. 3a) also confirms the absence of a bias possibly introduced by current heating of the transfer resistor up to 60 µA, at best. Testing much higher currents *via* measurements of $R_{H\text{-}G}$ according to our protocol is not possible mainly because of increasing errors caused by more important heating of this transfer resistor and deviation of $R_{H\text{-}GaAs}$ from $h/(2e^2)$ above $10^{-10}$. That is the reason why $R_{xx}$ measurements were preferred, as reported in the main text. The $R_{xx}$ data reported in Fig 5a and supporting the determination of $I_c$ presented in Fig 5b have been obtained with the magnetic field in the direction opposite to that of the magnetic field used to collect the data presented in Fig 3. We have checked the absence of significant effect of the magnetic field direction on the measurements of $R_{xx}$, which is an additional proof of the device quality.

### 6.3. Additional information about the dependence of $R_{H\text{-}G}$ on temperature.

For the study of the robustness of the Hall resistance quantization at higher temperatures, the Hall resistance measurements have been performed with the central transversal terminal-pair (B, F). In given conditions of magnetic field, temperature, and current, the measurements of $R_{xx}$ and $R_{H\text{-}G}$ have been performed in the two directions of the magnetic field. We report in Fig. S8a, the two sets of measurements, respectively performed with + $B$ and – $B$. For $R_{H\text{-}G}$, interestingly, there is no significant deviations from $R_{H\text{-}GaAs}$ that clearly changes its sign with the magnetic field direction. Note that in case of significant deviation of $R_{H\text{-}G}$, as observed at very low magnetic field and high current the sign of the deviation changes its sign with the magnetic field direction, as explained by the coupling mechanism of $R_{H\text{-}G}$ to $R_{xx}$. Moreover, $\Delta R_H/R_{H\text{-}odd}$ and $\Delta R_H/R_{H\text{-}even}$, as severe quantization criteria previously defined, are both zero within about $1\times10^{-9}$, as shown on Fig. S8c. This strongly supports that there is no deviation of $R_{H\text{-}G}$ from $R_{H\text{-}GaAs}$ in the conditions explored up to 10 K, within $1\times10^{-9}$. To account for this important result and enriching the data presented with the maximum information available, Fig 8a presents the average of the measurements performed with + $B$ and – $B$, that is to say $\Delta R_H/R_{H\text{-}even}$. The uncertainty associated is the maximum of the experimental standard deviation of the mean of the two measurements (+ $B$ and – $B$) and of the combined standard uncertainty calculated from the uncertainty of each of the two. The fact that $\Delta R_H/R_{H\text{-}odd}$ is zero within $1\times10^{-9}$, indicates that the averaging process (calculation of $\Delta R_H/R_{H\text{-}even}$) does not mask any significant deviations, in the limit of the uncertainty of $1\times10^{-9}$. The Fig 8b displays average value of the measurements of $R_{xx}$ performed with + $B$ and – $B$, the uncertainty is the combined standard uncertainty of the two measurements.

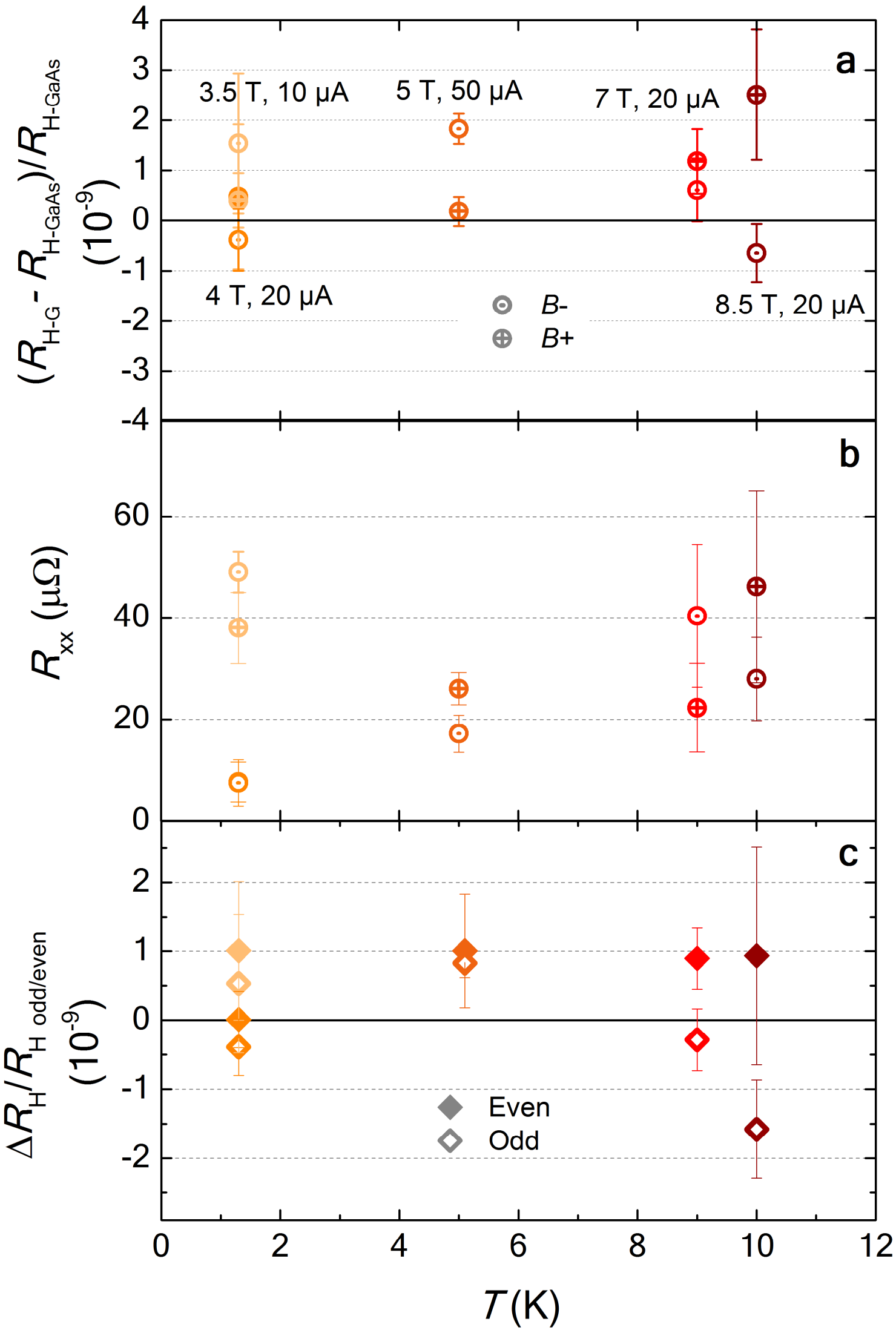

**Figure S8.** (a) Relative deviation of $R_{H-G}$ from $R_{H-GaAs}$ versus temperature, measured using the terminal-pairs (B,F) for both directions of the magnetic field. The operation magnetic field and current are also indicated. (b) Corresponding measurements of the longitudinal resistance per square (average of the four measurements along the device edges). (d) $\Delta R_H/R_H$ odd (empty symbols) and even (filled symbols) calculated with the data presented in (a).

### 6.4. Quantization accuracy tests at the lowest uncertainty.

With the objective to test the Hall quantization accuracy within the lowest uncertainty, we have repeated the $R_{H-G}$ measurements.

First, at 5 T, 1.3 K, 50 µA, eight measurements were carried out with the central pair (B, F), added to four measurements with each of the adjacent pairs (A, G) and (C, E) (half of each set being performed in one *B*-field direction and the second half with *B* in the opposite direction).

The histogram of all the sixteen $(\Delta R_H/R_H)_{-G}=(R_{H-G}-R_{H-GaAs})/R_{H-GaAs}$ values is presented in Figure 6a. The mean value is $-16\times10^{-11}$ and the experimental standard deviation $61\times10^{-11}$. The description of the histogram with a Gaussian calculated with this statistical data is rather good, considering the small number of measurements (Fig. 6a). This establishes that the noise limiting the precision of our measurement of $(\Delta R_H/R_H)_{-G}$ is uncorrelated random. This justifies the standard treatment of uncertainty and suggests that the result can be improved by repeating measurements. Since the data shows no sign of any systematic errors, the results obtained with combined standard uncertainty ranging from $32\times10^{-11}$ to $54\times10^{-11}$ can be combined to give a weighted mean of $(-11.4\pm10.2)\times10^{-11}$.
The observed accuracy of the quantization is strongly supported by the agreement between the weighted means of data measured with *B*+ and *B*- within $10.2\times10^{-11}$, expressed by $\Delta R_H/R_{H\text{-odd}}=(-1.87\pm10.2)\times10^{-11}$. $\Delta R_H/R_{H\text{-even}}$, the quantity previously reported, is equal to $(-11.4\pm10.2)\times10^{-11}$. This means that the maximum value for the dissipative longitudinal resistance would be 2 μΩ, compatible with the direct measurements of $R_{xx}$ that have been carried out.

To go beyond, we have refined the calculation of the Hall resistance quantization accuracy by considering all $R_{H\text{-G}}$ measurements performed in the best Hall quantization conditions (*B*, *T*, *I*, measurement configuration). The selection criteria for the measurement conditions were based on $\Delta R_H/R_{H\text{-odd}}=\Delta R_H/R_{H\text{-even}}<0.5\times10^{-9}$ *i.e.* $R_{H\text{-G}}$ invariance with the magnetic field direction, and also the dispersion between $R_{H\text{-G}}$ measurements below $1\times10^{-9}$. This led us to consider measurements performed at:

- (±5 T, 1.3 K, 50 μA), using the three transversal terminal-pairs: eight measurements with (B, F), four measurements with (A, G), four measurements with (C, E); half of each set in *B*+, half in *B*-; this is the set of measurements described previously,
- (±4 T, 1.3K, 20 μA), using the three transversal terminal-pairs: two measurements with (B, F), two measurements with (A, G), two measurements with (C, E); half of each set in *B*+, half in *B*-,
- (±6 T, 1.3K, 50 μA), using the central transversal terminal-pair (B, F) only: one measurement in *B*+, one measurement in *B*-,
- (±6 T, 1.3K, 60 μA), using the central transversal terminal-pair (B, F) only: one measurement in *B*+, one measurement in *B*-,
- (±8 T, 1.3K, 50 μA), using the central transversal terminal-pair (B, F) only: one measurement in *B*+, one measurement in *B*-.

The histogram of the twenty-eight selected values is presented in Figure 6a, it is pretty well described by a Gaussian calculated from their mean value of $0.1\times10^{-11}$ and their experimental standard deviation of $56\times10^{-11}$, which undoubtedly confirms that the $(\Delta R_H/R_H)_{\text{-G}}$ measurement noise is uncorrelated random. The weighted mean of the values is $(-0.9\pm8.2)\times10^{-11}$. The preservation of the accuracy when adding measurements at 4 T, 6 T, and 8 T, shows that $R_{H\text{-G}}$ quantization does not depend on *B*, in this magnetic field range, which is an additional proof of the Hall quantization robustness in this graphene device. As previously explained, the uncertainty components beyond random noise are negligible. The final uncertainty is limited by the white noise of the null detector measuring the balance of the voltages at the terminals of the two resistances compared in the resistance bridge. It could be further reduced by implementing the quantum Wheatstone bridge technique as used in Ref *(43)* where quantized Hall resistance comparisons were performed within record combined standard uncertainty of $3.2\times10^{-11}$.
Finally, because they are not correlated (different measurement protocols, differents experimental setups and instrumentation), it is possible to combine the result of our graphene vs. GaAs/AlGaAs ultimate comparison with the the result of the previous similar comparison by Janssen et al. *(25)*, in a weighted mean, in order to establish the QHE universality at $(-2.7\times10^{-11}\pm6.0\times10^{-11})$.